\documentclass[11pt,a4paper]{article}
\pdfoutput=1
\usepackage{jcappub}
\usepackage{ifthen}
\usepackage{epsfig}
\usepackage{booktabs} 
\usepackage{natbib}
\usepackage{bbold}

\newcommand{\beqra}{\begin{flalign}}
\newcommand{\eeqra}{\end{flalign}}
\newcommand{\beq}{\begin{equation}}
\newcommand{\eeq}{\end{equation}}
\newcommand{\mathd}{\mathrm{d}}
\usepackage{ccicons} 
\usepackage{listings}
\usepackage{color}   
\usepackage{braket}  
\usepackage{hyperref}
\usepackage{dsfont}  
\usepackage{leftidx} 
\usepackage{bm}
\usepackage{bbm} 
\usepackage{subfigure}
\allowdisplaybreaks

\title{WIMP capture by the Sun in the effective theory of dark matter self-interactions}
\author[a]{Riccardo Catena}
\author[a,b,c]{and Axel Widmark}
\affiliation[a]{Chalmers University of Technology, Department of Physics, SE-412 96 G\"oteborg, Sweden}
\affiliation[b]{The Oskar Klein Centre for Cosmoparticle Physics, AlbaNova, SE-106 91 Stockholm, Sweden}
\affiliation[c]{Department of Physics, Stockholm University, AlbaNova, SE-106 91 Stockholm, Sweden}
\emailAdd{catena@chalmers.se}
\emailAdd{waxel@student.chalmers.se}

\abstract{We study the capture of WIMP dark matter by the Sun in the non-relativistic effective theory of dark matter self-interactions.~The aim is to assess the impact of self-interactions on the expected neutrino flux from the annihilation of WIMPs trapped in the Sun in a model independent manner.~We consider all non-relativistic Galilean invariant self-interaction operators that can arise from the exchange of a heavy particle of spin less than or equal to 1 for WIMPs of spin equal to 0, 1/2 and 1.~We show that for interaction operators depending at most linearly on the momentum transfer, the WIMP-induced neutrino flux can be enhanced by several orders of magnitude compared to the same flux in absence of self-interactions.~This is true even for standard values of the thermally averaged annihilation cross-section.~This conclusion impacts the analysis of present and future observations performed at neutrino telescopes.}

\begin{document}
\maketitle

\section{Introduction}
Evidence for a dark matter component of the Universe has been gathered over a broad range of physical scales in the past decades~\cite{Jungman:1995df,Bergstrom00}.~In the paradigm of Weakly Interacting Massive Particles (WIMPs) as a dark matter candidate, WIMPs are expected to interact with nuclei via elastic scattering~\cite{Goodman:1984dc}.~This property of WIMPs has a variety of implications as far as dark matter particle detection is concerned~\cite{Bertone:2004pz}.~In particular, WIMPs could scatter to gravitationally bound orbits while crossing the solar interior, accumulate at the Sun's centre, and finally annihilate into Standard Model particles at an observable rate~\cite{Silk:1985ax}.~In this scenario, the physical observable is the flux of energetic neutrinos produced in the decay chain of WIMP primary annihilation products.~Neutrino telescopes such as IceCube, Super-Kamiokande, ANTARES and Baikal are currently searching for this as of yet hypothetical neutrino signal~\cite{Aartsen:2016exj,Choi:2015ara,Adrian-Martinez:2016gti,Avrorin:2014swy}.

Progress in understanding the general properties of the capture of WIMPs by the Sun has recently been made through the use of effective theories~\cite{Catena:2015iea,Blumenthal:2014cwa,Liang:2013dsa,Guo:2013ypa}.~In particular, the non-relativistic effective theory of dark matter-nucleon interactions has allowed us to gain new insights on the actual complexity of the capture process~\cite{Catena:2015iea,Catena:2015uha}.~The effective theory of dark matter-nucleon interactions has been proposed in the context of dark matter direct detection~\cite{Chang:2009yt,Fan:2010gt,Fitzpatrick:2012ix,Fitzpatrick:2012ib}, and developed in~\cite{Fornengo:2011sz,Menendez:2012tm,Cirigliano:2012pq,Anand:2013yka,DelNobile:2013sia,Klos:2013rwa,Peter:2013aha,Hill:2013hoa,Catena:2014uqa,Catena:2014hla,Catena:2014epa,Gluscevic:2014vga,Panci:2014gga,Vietze:2014vsa,Barello:2014uda,Catena:2015uua,Schneck:2015eqa,Dent:2015zpa,Catena:2015vpa,Kavanagh:2015jma,D'Eramo:2016atc,Catena:2016hoj,Kahlhoefer:2016eds}.~It allows to compute elastic and inelastic WIMP-nucleus scattering cross-sections without assuming a specific coupling of dark matter to quarks or gluons.~Its application to the study of WIMP capture by the Sun is based upon the numerical shell-model calculations performed in~\cite{Catena:2015uha}, where all nuclear response functions predicted by the theory are computed and made publicly available for the most abundant elements in the Sun.~A variety of new phenomena have been identified pursuing this approach, both concerning the complementarity of direct detection experiments and the search for WIMPs at neutrino telescopes, and regarding the potential of neutrino telescopes themselves~\cite{Catena:2015iea,Catena:2015uha}.

WIMPs can also be captured by the Sun via self-interaction, as shown in~\cite{Zentner:2009is}.~WIMPs self-interactions are particularly interesting in the context of dark matter searches at neutrino telescopes, since they can enhance the rate of WIMP capture by the Sun, and therefore amplify the associated WIMP-induced neutrino flux~\cite{Albuquerque:2013xna,Chen:2014oaa,Chen:2015uha}.~So far, the use of effective theory methods has been restricted to the capture of WIMPs via scattering by nuclei.~In this paper we perform the first calculation of the rate of dark matter capture by the Sun in the effective theory of short-range dark matter self-interactions, recently proposed in~\cite{Bellazzini:2013foa}.~In contrast to previous findings~\cite{Zentner:2009is}, we show that in the general effective theory of dark matter self-interactions large neutrino signal amplifications are expected, even for standard thermally averaged annihilation cross-sections and momentum/velocity independent dark matter self-interactions.

The paper is organised as follows.~In Sec.~\ref{capture} we review the capture of WIMPs in the Sun via elastic scattering by nuclei and introduce the equations describing WIMPs capture via self-interaction.~Sec.~\ref{sec:theory} is devoted to the effective theory of dark matter-nucleon interactions and the effective theory of dark matter self-interactions.~The two theories will be used to calculate scattering and capture rates in the Sun in the subsequent section.~We present our results for the rate of dark matter capture by the Sun in the effective theory of dark matter self-interactions in Sec.~\ref{sec:results}.~We highlight the broad applicability of our results and conclude in Sec.~\ref{sec:conclusions}.~Key equations for the calculation of WIMP scattering and self-scattering cross-sections are in Appendix~\ref{sec:appDM}.

\section{Dark matter capture by the Sun revisited}
\label{capture}
In this section we review the various processes that determine the number of dark matter particles trapped in the Sun at a given time.
\subsection{Capture}

WIMP dark matter particles travelling through the Sun are expected to interact with the solar interior, scattering to gravitationally bound orbits.~In this scenario, WIMPs accumulate at the Sun's centre where they eventually annihilate producing an observable flux of energetic neutrinos.~This accumulation process is known as dark matter capture.~In the standard paradigm of WIMP dark matter, capture occurs via scattering by nuclei in the Sun.~The rate of scattering from a velocity $w$ to a velocity less than the local escape velocity $v(r)$ at a solar radius $r$ is given by~\cite{Gould:1987ir}
\begin{equation}
\Omega_{v}^{c\,-}(w)= \sum_T n_T w\,\Theta\left( \frac{\mu_T}{\mu^2_{+,T}} - \frac{u^2}{w^2} \right)\int_{E u^2/w^2}^{E \mu_T/\mu_{+,T}^2} {\rm d}E_r\,\frac{{\rm d}\sigma_{\chi T}\left(E_r,w^2\right)}{{\rm d}E_r}\,,
\label{eq:omega}
\end{equation}
where $E=m_\chi w^2/2$, $m_\chi$ is the dark matter mass, $w=\sqrt{u^2+v(r)^2}$ and $u$ are, respectively, the dark matter particle velocity at the scattering point and at infinity, and $d\sigma_{\chi T}/dE_r$ is the differential cross-section for dark matter scattering by nuclei of mass $m_T$ and density in the Sun $n_T(r)$.~The sum in Eq.~(\ref{eq:omega}) extends over the most abundant elements in the Sun, and the dimensionless parameters $\mu_T$ and $\mu_{\pm,T}$  are defined as follows:~$\mu_T\equiv m_\chi/m_T$ and $\mu_{\pm,T}\equiv (\mu_T\pm1)/2$.~The integration is performed over all kinematically allowed recoil energies $E_r$.~While in computing $\Omega_{v}^{c\,-}(w)$ the most common approach is to assume a constant total cross-section $\sigma_{\chi T}$, here we will consider a more general class of dark matter interactions, as we will see in Sec.~\ref{sec:theory}.

The differential capture rate per unit volume is then obtained from Eq.~(\ref{eq:omega}).~It reads as follows~\cite{Gould:1987ir} 
\begin{equation}\label{dCdV}
    \frac{\mathd C_c}{\mathd V} = \int_0^\infty \mathd u\frac{f(u)}{u}w\Omega_v^{c\,-}(w) \,,
\end{equation}
where $f(u)$ is the WIMP speed distribution at infinity boosted in the Sun's rest frame.~For the latter we adopt a Maxwell-Boltzmann distribution and assume a Local Standard of Rest of 220 km/s.~Finally, the total rate of dark matter capture via scattering by nuclei in the Sun is given by:
\begin{equation}\label{C}
    C_c =\int_0^{R_\odot} \mathd r\,4\pi r^2 \frac{\mathd C_c}{\mathd V} \,,
\end{equation}
where $R_\odot$ is the radius of the Sun, and spherical symmetry is assumed performing the volume integral. 

As pointed out in~\cite{Zentner:2009is}, WIMP self-interactions provide a second intriguing mechanism for dark matter capture by the Sun.~In the case of capture via dark matter self-interaction, the capture rate is computed as above.~Since incoming and target particle are identical, Eq.~(\ref{dCdV}) now reads as follows
\begin{equation}\label{selfcapturesigma}
    \Omega_v^{s\,-}(w) =
    \epsilon w\,\Theta\left(1-\frac{u^2}{w^2}\right)
    \int_{E u^2/w^2}^{E}\mathd E_r\frac{\mathd \sigma_{\chi \chi}(E_r,w^2)}{\mathd E_r},
\end{equation}
where $\epsilon(r)$ is the number density of already trapped dark matter particles, and $\mathd \sigma_{\chi \chi}/\mathd E_r$ is the differential cross-section for dark matter self-scattering.~We will compute the latter as explained in Sec.~\ref{sec:theory}.~By replacing $\Omega^{c\,-}_v$ with $\Omega^{s\,-}_v$ in Eqs.~\eqref{dCdV} and \eqref{C}, we obtain the differential and total rate of dark matter capture by the Sun via self-interaction; $\mathd C_s/\mathd V$ and $C_s$, respectively.~We denote by $\Gamma_s=N_\chi C_s$ the number of WIMPs captured via dark matter self-interaction per unit time.~$N_\chi(t)$ is the total number of WIMPs already trapped in the Sun at the time $t$. 

The gravitationally bound dark matter particles are assumed to quickly thermalise, regardless of the details of the capture process.~Their radial distribution then follows a thermal profile given by
\begin{equation}\label{thermalprofile}
\epsilon(r) \propto \text{exp}\left[-\frac{m_\chi\phi(r)}{T_c}\right],    
\end{equation}
where $\phi(r)$ is the total gravitational potential at $r$ and $T_c=1.57\times10^7$~K is the solar core temperature.~Although this approximation is expected to be valid for most of the parameter values considered here, only detailed numerical calculations can determine the actual distribution and thermalisation time of WIMPs in the Sun~\cite{Vincent:2013lua}.~Following~\cite{Gondolo:2004sc}, we model the gravitational potential $\phi(r)$ from the radial density and mass fractions of the elements in the Sun.~We consider the sixteen elements:~H, $^3$He, $^4$He, $^{12}$C, $^{14}$N, $^{16}$O, $^{20}$Ne, $^{23}$Na, $^{24}$Mg, $^{27}$Al, $^{28}$Si, $^{32}$S, $^{40}$Ar, $^{40}$Ca, $^{56}$Fe, and $^{59}$Ni, with radial density and mass fractions as given in~\cite{massf} and \cite{raddens}, respectively.

\subsection{Annihilation}
The increase in number of WIMPs due to capture via scattering by nuclei or self-scattering is contrasted by WIMP annihilation in the Sun.~The average number of WIMP annihilations per unit time, $\Gamma_a$, is given by
\begin{equation}
\Gamma_a=\frac{1}{2}\int d^3{\bf x} \,\epsilon^2({\bf x}) \,\langle \sigma_{\rm ann} v_{\rm rel} \rangle \,,
\label{eq:Gamma}
\end{equation}
where $\langle \sigma_{\rm ann} v_{\rm rel} \rangle$ is the thermally averaged dark matter annihilation cross-section times the relative velocity $v_{\rm rel}$, and ${\bf x}$ denotes the three-dimensional WIMP position vector.~Eq.~(\ref{eq:Gamma}) implies the relation $\Gamma_a=C_a N_\chi^2/2$, where we have introduced the annihilation rate $C_a$, which is given by 
\begin{equation}
C_a = \langle \sigma_{\rm ann} v_{\rm rel} \rangle \frac{V_2}{V_1^2}\,.
\label{eq:CA}
\end{equation}
In this expression
\begin{equation}
\label{eq:vol}
V_1 = \int d^3{\bf x} \,\frac{\epsilon({\bf x})}{\epsilon_0} \,; \qquad \qquad V_2 = \int d^3{\bf x} \,\frac{\epsilon^2({\bf x})}{\epsilon_0^2}\,,
\end{equation}
and $\epsilon_0$ is the dark matter density at the centre of the Sun.~Starting from Eq.~(\ref{thermalprofile}), the effective volumes $V_1$ and $V_2$ can be calculated numerically, or analytically under the assumption of a constant solar density.~They represent a tiny fraction of the total volume of the Sun, which implies that most of the solar WIMPs are concentrated at the Sun's centre after capture and thermalisation.

The differential neutrino flux produced by WIMP annihilation in the Sun is proportional to the rate $\Gamma_a$~\cite{Jungman:1995df}:
\begin{equation}
    \frac{\mathd \Phi_\nu}{\mathd E_\nu} =
    \frac{\Gamma_a}{4\pi D^2}\sum_f B^f_\chi \frac{\mathd N^f_\nu}{\mathd E_\nu}\,.
\label{eq:nuflux}
\end{equation}
In Eq.~(\ref{eq:nuflux}), $E_\nu$ is the neutrino energy, $D$ is the observer's distance from the Sun, $B^f_\chi$ is the branching ratio for dark matter pair annihilation into the final state $f$, and $\mathd N^f_\nu / \mathd E_\nu$ is the neutrino energy spectrum at the detector produced by the decay chain induced by the final state $f$.

\subsection{Equilibration}

\begin{figure}
  \begin{center}
    \includegraphics[width=0.7\textwidth]{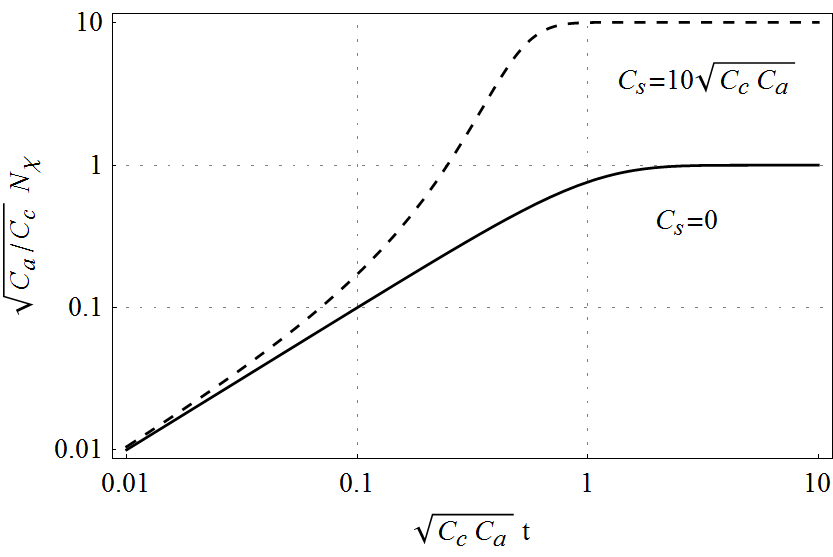}
  \end{center}
  \caption{Total number of trapped dark matter particles, $N_\chi$, as a function of time $t$.~The solid line represents the case of no self-interaction, $C_s=0$, whereas the dashed line corresponds to the case of strong self-interaction:~$C_s=10\sqrt{C_cC_a}$.}
  \label{Nplot}
\end{figure}

The number of dark matter particles that are captured in the Sun's core, $N_\chi$, obeys the differential equation \cite{Zentner:2009is}

\begin{align}\label{diffeqwithself}
    \frac{\mathd N_\chi}{\mathd t}=C_c+C_sN_\chi-C_aN_\chi^2,
\end{align}
where $C_c$ is the capture rate by nuclei, $C_s$ is the capture rate via self-interaction, and $C_a$ is the annihilation rate. It has solution

\begin{equation}
    N_\chi = \frac{ C_c\tanh(t/\zeta) }{\zeta^{-1} -C_s \tanh(t/\zeta)/2}\,,
\end{equation}
where
\begin{align}
    \zeta=\frac{1}{\sqrt{C_cC_a+C_s^2/4}}\,.
\end{align}
Fig.~\ref{Nplot} shows this solution as a function of time $t$. As seen in this figure, $N_\chi$ will eventually tend to an equilibrium configuration, for which the total rate of capture and rate of annihilation are equal.

\noindent A kew quantity in our analysis is the time-dependent ratio $\beta(t)$:
\begin{align}\label{beta}
    \beta(t)
    &\equiv \left(\frac{N_{C_s\neq0}(t)}{N_{C_s=0}(t)}\right)^2 \nonumber\\
    &= \left(\frac{\sqrt{C_cC_a}\coth(\sqrt{C_cC_a}t)\tanh(t/\zeta)}{\zeta^{-1}-C_s\tanh(t/\zeta)/2}
    \right)^2\,,
\end{align}
where $N_{C_s=0}$ $(N_{C_s\neq0})$ is $N_\chi$ computed assuming $C_s=0$ ($C_s\neq0$).~It measures the relative enhancement of $\Gamma_a$, and therefore of the WIMP-induced neutrino flux, due to dark matter self-interactions.

Notably, the ratio $\beta(t)$ does not depend on $C_c$ and $C_a$ separately.~It only depends on their product $C_c C_a$.~This means that $\beta(t)$ remains unchanged when $C_c$ and $C_a$ are, respectively, increased and decreased by the same factor.~This key observation appears to be in contrast with the numerical results, and specifically wiht Fig.~2, in Ref.~\cite{Zentner:2009is}.~At equilibrium, as $t\rightarrow\infty$, $\beta(t)$ becomes
\begin{equation}\label{betaequilibriumsolution}
    \beta_\text{eq} = \frac{C_s^2}{2C_cC_a}+\sqrt{\left(\frac{C_s^2}{2C_cC_a}\right)^2+1}.
\end{equation}
In the case of dark matter self-capture being the dominant process, the amplification factor takes the approximate form $\beta_\text{eq}\simeq C_s^2/(C_cC_a)$. In the opposite limit, where self-capture is weak, the amplification tends to $\beta_\text{eq}\simeq 1+C_s^2/(2C_cC_a)$, where the last term of this expression is small relative to 1.

We now focus on Eq.~(\ref{diffeqwithself}) in the limit $t\ll t_{\rm eq}$, where $t_{\rm eq}\equiv 1/\sqrt{C_c C_a}$ is an approximate equilibration time.~For $t\ll t_{\rm eq}$, $\epsilon(r)$ is small enough to neglect the rate $\Gamma_a$ in Eq.~(\ref{diffeqwithself}).~In this case we find the solution
\begin{equation}
    N_\chi(t \ll t_\text{eq})\simeq\frac{C_c}{C_s}(e^{C_st}-1)\,,
\end{equation}
and 
\begin{equation}\label{betanonequilibriumsolution}
    \beta(t \ll t_\text{eq}) \simeq \left(\frac{e^{C_st}-1}{C_st}\right)^2\,.
\end{equation}
For $t\ll t_{\rm eq}$, $\beta(t)$ does not depend on $C_c$.~However, while $C_c$ does not affect the \emph{relative} amplification, it very much affects the absolute rate of capture, the absolute flux of neutrinos and the value of $t_{\rm eq}$.

\section{Dark matter non-relativistic scattering in effective theories}
\label{sec:theory}
In this section we review the effective theory of dark matter-nucleon interactions~\cite{Fitzpatrick:2012ix} and the effective theory of dark matter self-interactions~\cite{Bellazzini:2013foa}.~The two theories are used to calculate the differential cross-sections for dark matter-nucleus scattering and dark matter self-scattering appearing in Eqs.~(\ref{eq:omega}) and (\ref{selfcapturesigma}), respectively.~Constraints on the parameter space of the two theories are also discussed.~For further details, we refer to the original literature on the topic~\cite{Chang:2009yt,Fan:2010gt,Fornengo:2011sz,Fitzpatrick:2012ix,Fitzpatrick:2012ib,Menendez:2012tm,Cirigliano:2012pq,Anand:2013yka,DelNobile:2013sia,Klos:2013rwa,Peter:2013aha,Hill:2013hoa,Catena:2014uqa,Catena:2014hla,Catena:2014epa,Gluscevic:2014vga,Panci:2014gga,Vietze:2014vsa,Barello:2014uda,Catena:2015uua,Schneck:2015eqa,Bellazzini:2013foa}.

\begin{table}[t]
    \centering
    \begin{tabular}{ll}
    \toprule
        $\hat{\mathcal{O}}_1 = \mathbb{1}_{\chi N}$ & $\hat{\mathcal{O}}_9 = i{\bf{\hat{S}}}_\chi\cdot\left({\bf{\hat{S}}}_N\times\frac{{\bf{\hat{q}}}}{m_N}\right)$  \\
        $\hat{\mathcal{O}}_3 = i{\bf{\hat{S}}}_N\cdot\left(\frac{{\bf{\hat{q}}}}{m_N}\times{\bf{\hat{v}}}^{\perp}\right)$ \hspace{2 cm} &   $\hat{\mathcal{O}}_{10} = i{\bf{\hat{S}}}_N\cdot\frac{{\bf{\hat{q}}}}{m_N}$   \\
        $\hat{\mathcal{O}}_4 = {\bf{\hat{S}}}_{\chi}\cdot {\bf{\hat{S}}}_{N}$ &   $\hat{\mathcal{O}}_{11} = i{\bf{\hat{S}}}_\chi\cdot\frac{{\bf{\hat{q}}}}{m_N}$   \\                                                                             
        $\hat{\mathcal{O}}_5 = i{\bf{\hat{S}}}_\chi\cdot\left(\frac{{\bf{\hat{q}}}}{m_N}\times{\bf{\hat{v}}}^{\perp}\right)$ &  $\hat{\mathcal{O}}_{12} = {\bf{\hat{S}}}_{\chi}\cdot \left({\bf{\hat{S}}}_{N} \times{\bf{\hat{v}}}^{\perp} \right)$ \\                                                                                                                 
        $\hat{\mathcal{O}}_6 = \left({\bf{\hat{S}}}_\chi\cdot\frac{{\bf{\hat{q}}}}{m_N}\right) \left({\bf{\hat{S}}}_N\cdot\frac{\hat{{\bf{q}}}}{m_N}\right)$ &  $\hat{\mathcal{O}}_{13} =i \left({\bf{\hat{S}}}_{\chi}\cdot {\bf{\hat{v}}}^{\perp}\right)\left({\bf{\hat{S}}}_{N}\cdot \frac{{\bf{\hat{q}}}}{m_N}\right)$ \\   
        $\hat{\mathcal{O}}_7 = {\bf{\hat{S}}}_{N}\cdot {\bf{\hat{v}}}^{\perp}$ &  $\hat{\mathcal{O}}_{14} = i\left({\bf{\hat{S}}}_{\chi}\cdot \frac{{\bf{\hat{q}}}}{m_N}\right)\left({\bf{\hat{S}}}_{N}\cdot {\bf{\hat{v}}}^{\perp}\right)$  \\
        $\hat{\mathcal{O}}_8 = {\bf{\hat{S}}}_{\chi}\cdot {\bf{\hat{v}}}^{\perp}$  & $\hat{\mathcal{O}}_{15} = -\left({\bf{\hat{S}}}_{\chi}\cdot \frac{{\bf{\hat{q}}}}{m_N}\right)\left[ \left({\bf{\hat{S}}}_{N}\times {\bf{\hat{v}}}^{\perp} \right) \cdot \frac{{\bf{\hat{q}}}}{m_N}\right] $ \\                                                                               
    \bottomrule
    \end{tabular}
    \caption{Interaction operators that appear in Eqs.~(\ref{eq:H_chiT}) and (\ref{eq:H_chichi}).~Here $m_N$ is the nucleon mass and all operators have the same mass dimension.~We omit the nucleon index $i$ which is only relevant for dark matter-nucleon interactions.} 
    \label{tab:operators}
\end{table}

\subsection{Symmetries and interaction operators}
The transition amplitude for non-relativistic scattering of dark matter by nuclei, or for dark matter self-scattering, is restricted by momentum conservation and Galilean invariance, i.e.~the invariance under constant shifts of particle velocities.~These restrictions imply that in the non-relativistic limit any given quantum mechanical interaction operator describing dark matter-nucleon interactions or dark matter self-interactions can be expressed in terms of the following building blocks: ${\bf{\hat{q}}}$, ${\bf{\hat{v}}}^{\perp}$, ${\bf{\hat{S}}}_\chi$ and ${\bf{\hat{S}}}_N$, representing the momentum transfer operator, the transverse relative velocity operator, and the dark matter and $N$ particle spin operators, respectively.~We denote by $N$ the second particle participating in the scattering, which can be either a second dark matter particle or a target nucleon, depending on the cross-section in analysis.

Let us denote by $\hat{\mathcal{H}}$ the interaction Hamiltonian underlying the scattering process.~Without further restrictions, $\hat{\mathcal{H}}$ will include an infinite number of interaction operators:~all scalar combinations of ${\bf{\hat{q}}}$, ${\bf{\hat{v}}}^{\perp}$, ${\bf{\hat{S}}}_\chi$ and ${\bf{\hat{S}}}_N$.~However, $\hat{\mathcal{H}}$ can be simplified when the momentum transferred in the WIMP scattering (or self-scattering) is small compared to the mass of the particle that mediates the interaction.~In this case, $\hat{\mathcal{H}}$ can be expanded in powers of ${\bf{\hat{q}}}$.~Truncating the expansion at second order, but keeping one term of the order of $|{\bf \hat{q}}|^2{\bf \hat{v}}^{\perp} $~\cite{Anand:2013yka}, only eighteen Galilean invariant interaction operators can be generated~\cite{Dent:2015zpa}.~In this analysis we use the notation introduced in~\cite{Anand:2013yka} to characterise different dark matter-nucleon and dark matter self-interaction operators.~Among the eighteen interaction operators, we neglect:~$\hat{\mathcal{O}_{2}}$, which is quadratic in ${\bf{\hat{v}}}^{\perp}$; $\hat{\mathcal{O}}_{16}$, which is a linear combination of $\hat{\mathcal{O}}_{12}$ and $\hat{\mathcal{O}}_{15}$; and $\hat{\mathcal{O}}_{17}$ ($\hat{\mathcal{O}}_{18}$), which is degenerate with $\hat{\mathcal{O}}_{5}$ (a linear combination of $\hat{\mathcal{O}}_{9}$ and $\hat{\mathcal{O}}_{10}$), if operator interference can be neglected, as we do here.~In Tab.~\ref{tab:operators} we list all interaction operators considered in this investigation. 

From the operators in Tab.~\ref{tab:operators}, one can construct general Hamiltonian densities for dark matter-nucleon and dark matter self-interactions.~Under the assumption of one-body dark matter-nucleon interactions, the most general Hamiltonian density for dark matter-nucleus scattering is given by
\begin{equation}
\hat{\mathcal{H}}_{\chi T}= \sum_{i=1}^{A}  \sum_{\tau=0,1} \sum_{k} c_k^{\tau}\hat{\mathcal{O}}_{k}^{(i)} \, t^{\tau}_{(i)} \,.
\label{eq:H_chiT}
\end{equation}
The operators $t^0_{(i)}=\mathbb{1}_{2\times 2}$ and $t^1_{(i)}=\tau_3$, where $\tau_3$ is the third Pauli matrix, are defined in the isospin space of the $i$-th nucleon, and $A$ is the mass number of the target nucleus.~The isoscalar and isovector coupling constants $c_k^0$ and $c_k^1$, respectively, have dimension mass to the power $-2$, and are related to the coupling constants for protons and neutrons as follows:~$c^{p}_k=(c^{0}_k+c^{1}_k)/2$, and $c^{n}_k=(c^{0}_k-c^{1}_k)/2$.

The most general Hamiltonian density for dark matter self-interactions $\hat{\mathcal{H}}_{\chi\chi}$ is obtained by removing $t^\tau_{(i)}$ and the sum over the $\tau$ and $i$ indexes in Eq.~(\ref{eq:H_chiT}).~Furthermore, in the case of dark matter self-interactions there is no distinction between isoscalar and isovector couplings, and we will therefore denote the corresponding coupling constants by $c_k$, with no upper index:
\begin{equation}
\hat{\mathcal{H}}_{\chi \chi}= \sum_{k} c_k \, \hat{\mathcal{O}}_{k} \,.
\label{eq:H_chichi}
\end{equation}
Neglecting operator interference, Eqs.~(\ref{eq:H_chiT}) and (\ref{eq:H_chichi}) contain all non-relativistic Galilean invariant operators that can arise from the exchange of a heavy particle of spin $\le 1$ for WIMPs of spin equal to 0, 1/2 and 1.

\subsection{Scattering cross-sections}
Given the Hamiltonian density $\hat{\mathcal{H}}_{\chi T}$ in Eq.~(\ref{eq:H_chiT}), the transition probability for dark matter-nucleus scattering is proportional to
\begin{align}
\langle |\mathcal{M}|^2\rangle_{\chi T} =  \frac{4\pi}{2J_T+1}\sum_{\tau,\tau'} &\bigg[ \sum_{k=M,\Sigma',\Sigma''} R^{\tau\tau'}_k\left(v_T^{\perp 2}, {q^2 \over m_N^2} \right) W_k^{\tau\tau'}(q^2) \nonumber\\
&+{q^{2} \over m_N^2} \sum_{k=\Phi'', \Phi'' M, \tilde{\Phi}', \Delta, \Delta \Sigma'} R^{\tau\tau'}_k\left(v_T^{\perp 2}, {q^2 \over m_N^2}\right) W_k^{\tau\tau'}(q^2) \bigg] \,, \nonumber\\
\label{eq:M} 
\end{align}
where $\mathcal{M}$ is the scattering amplitude normalised as in~\cite{Anand:2013yka}, $J_T$ and $m_T$ are the nuclear spin and mass, respectively, and angle brackets denote a sum (average) over the final (initial) spin-configurations.~The eight nuclear response functions $W_k^{\tau\tau'}$ in Eq.~(\ref{eq:M}) were introduced in~\cite{Fitzpatrick:2012ix}.~They are quadratic in matrix elements of nuclear charges and currents.~The eight functions $R^{\tau\tau'}_k$ in Eq.~(\ref{eq:M}) are known analytically, and depend on $q^2/m_N^2$ and $v_T^{\perp 2}=w^2-q^2/(4\mu_T^2)$, where $w$ is the dark matter-nucleus relative velocity, $m_N$ is the nucleon mass, and $\mu_T$ the reduced dark matter-nucleus mass.~The $R^{\tau\tau'}_k$ functions relevant for this analysis are in Appendix~\ref{sec:appDM}.~We use the $W_k^{\tau\tau'}$ functions computed in~\cite{Catena:2015uha} for the 16 most abundant elements in the Sun.~The differential cross-section for dark matter scattering by nuclei of mass $m_T$ can thus be written as follows
\begin{equation}
\frac{{\rm d}\sigma_{\chi T}(q^2,w^2)}{{\rm d}q^2} = \frac{1}{4\pi w^2} \, \langle |\mathcal{M}|^2\rangle_{\chi T} \,,
\label{eq:sigma}
\end{equation}
which is in general a function of the momentum transfer, and of the dark matter-nucleus relative velocity.

Analogously, from the Hamiltonian $\hat{\mathcal{H}}_{\chi \chi}$ we obtain the differential cross-section for dark matter self-scattering:
\begin{equation}
\frac{{\rm d}\sigma_{\chi \chi}(q^2,w^2)}{{\rm d}q^2} = \frac{1}{4\pi w^2} \, \langle |\mathcal{M}|^2\rangle_{\chi \chi} \,,
\label{eq:sigma2}
\end{equation}
where
\begin{align}
\langle |\mathcal{M}|^2\rangle_{\chi \chi} =  \frac{4\pi}{2J_\chi+1} \sum_{\,k=M,\Sigma',\Sigma''} R_k\left(v_T^{\perp 2}, {q^2 \over m_N^2} \right) W_k  \,, 
\label{eq:M2} 
\end{align}
$J_\chi$ is the dark matter particle spin, and $W_M=1/(8\pi)$, $W_{\Sigma'}=1/(4\pi)$ and $W_{\Sigma''}=1/(8\pi)$.~Assuming that dark matter is a point-like particle, all other response functions are zero.~Compared to Eq.~(\ref{eq:sigma}), the sum over the $\tau$ and $\tau'$ indexes has now consistently been dropped.~For definiteness, we will assume $J_\chi=1/2$.~From the relations $q^2=2 m_T E_r$ or $q^2=2 m_\chi E_r$, one can finally obtain the differential cross-sections ${\rm d}\sigma_{\chi T}/{\rm d} E_r=2 m_T \, {\rm d} \sigma_{\chi T}/{\rm d} q^2$ and ${\rm d}\sigma_{\chi \chi}/{\rm d} E_r=2 m_\chi \, {\rm d} \sigma_{\chi \chi}/{\rm d} q^2$, respectively.

\subsection{Constraints}
The coupling constants for dark matter-nucleon interactions $c_k^\tau$  are constrained by direct detection experiments and neutrino telescopes.~The strongest limits come from LUX in a wide range of $m_\chi$~\cite{Akerib:2016vxi}, for the majority of the operators~\cite{Catena:2014uqa}.~However, neutrino telescopes~\cite{Aartsen:2016exj} together with PICO-60~\cite{Amole:2015pla} and PICO-2L~\cite{Amole:2016pye} place stronger limits on the dark matter coupling to protons for the interaction operator $\hat{\mathcal{O}}_4$.~Neutrino telescopes are superior to LUX also concerning the limits on the strength of the $\hat{\mathcal{O}}_7$ interaction~\cite{Catena:2015iea}.~At the same time, CDMSlite~\cite{Agnese:2015nto}, SuperCDMS~\cite{Schneck:2015eqa} and CRESST-II~\cite{Angloher:2015ewa} set the most stringent limits on the coupling constants for dark matter-nucleon interactions for $m_\chi\lesssim 5$~GeV.~In this study we will use the upper limits on $c_k^\tau$ derived in~\cite{Catena:2015iea} from LUX.~Current LUX limits on $c_1^0$ are about a factor of 2 stronger~\cite{Akerib:2016vxi}.

We consider two upper limits for the self-scattering cross-section $\sigma_{\chi\chi}$, which is obtained by integrating Eq.~(\ref{eq:sigma2}) over all kinematically allowed recoil energies.~The first one arises from N-body simulations~\cite{Rocha:2012jg}
\begin{equation}\label{v=1000limit}
    \sigma_{\chi\chi} < 0.1\;\left(\frac{m_\chi}{\text{g}}\right)\;\text{cm}^2 = 1.78\times10^{-25}\; \left(\frac{m_\chi}{\text{GeV}}\right)\;\text{cm}^2\,,
\end{equation}
the second one from an analysis of the Bullet cluster~\cite{BulletClusterLimit08}:
\begin{equation}
    \sigma_{\chi\chi} < 1.25\;\left(\frac{m_\chi}{\text{g}}\right)\;\text{cm}^2 =2.23\times10^{-24}\; \left(\frac{m_\chi}{\text{GeV}}\right)\;\text{cm}^2\,.
\label{eq:bullet}
\end{equation}
Strictly speaking, the constraints in Eqs.~(\ref{v=1000limit}) and (\ref{eq:bullet}) apply to $\sigma_{\chi\chi}$ at the typical collisional velocity of the WIMPs forming the astrophysical system from which the limit has been derived.~In the former case we assume $w=1000$~km/s, in the latter $w=4700$~km/s, which is the inferred relative velocity of the two sub-clusters merging in the Bullet cluster.~Notice that only interaction operators different from  $\hat{\mathcal{O}}_1$ and  $\hat{\mathcal{O}}_4$ generate velocity dependent dark matter self-scattering cross-sections.~The upper bounds in Eqs.~(\ref{v=1000limit}) and (\ref{eq:bullet}) constrain the self-scattering cross-section $\sigma_{\chi\chi}$ at different velocities, placing complementary limits on the interaction operators in Tab.~\ref{tab:operators}.

\section{Results}
\label{sec:results}

\subsection{Analytic results}
We can accurately calculate the rate of capture via self-interaction, $C_s$, for the case of constant cross-section $\sigma_{\chi\chi}$, if we approximate the trapped WIMPs to be situated in the very centre of the Sun.~This is very close to the actual case, as the thermalised WIMPs are localised to less than a percent of the Sun's radius, and especially so in the high end of the mass spectrum that we consider. Using this assumption and a Maxwell-Boltzmann velocity distribution, the integrals are simple to perform analytically, and the capture rate via self-interaction takes the form~\cite{Zentner:2009is}
\begin{equation}
C_s = \sqrt{\frac{3}{2}} n_\chi \sigma_{\chi\chi} \frac{v^2(0)}{\bar{v}} \frac{\text{erf}(\eta)}{\eta},
\end{equation}
where $n_\chi=\rho_\chi/m_\chi$ is the local WIMP number density, $\rho_\chi=0.4$~GeV/cm$^3$~\cite{Catena:2009mf}, $v(0)$ is the escape velocity at the Sun's centre, and $\eta=\sqrt{3/2}v_\odot/\bar{v}$, where $\bar{v}=270$~km/s is the WIMP velocity dispersion~\cite{Catena:2011kv,Bozorgnia:2013pua} and $v_\odot=220$~km/s is the Local Standard of Rest velocity.~Using a WIMP mass of 100 GeV and a cross-section of $\sigma_{\chi\chi} = 10^{-23}$ cm$^2$, which is within current limits, we find
\begin{equation}
C_s = 2.76 \times 10^{-17}~{\rm s}^{-1}.
\end{equation}
The largest values for $\beta$ are found in the non-equilibrium region. Substituting this value of $C_s$ into Eq.~(\ref{betanonequilibriumsolution}) and using the current age of the Sun, $t_\odot=1.44\times 10^{17}$ s, we get the value $\beta(t_\odot)\simeq173$, which is significantly larger than 1.

\begin{figure}
  \begin{center}
    \includegraphics[width=\textwidth]{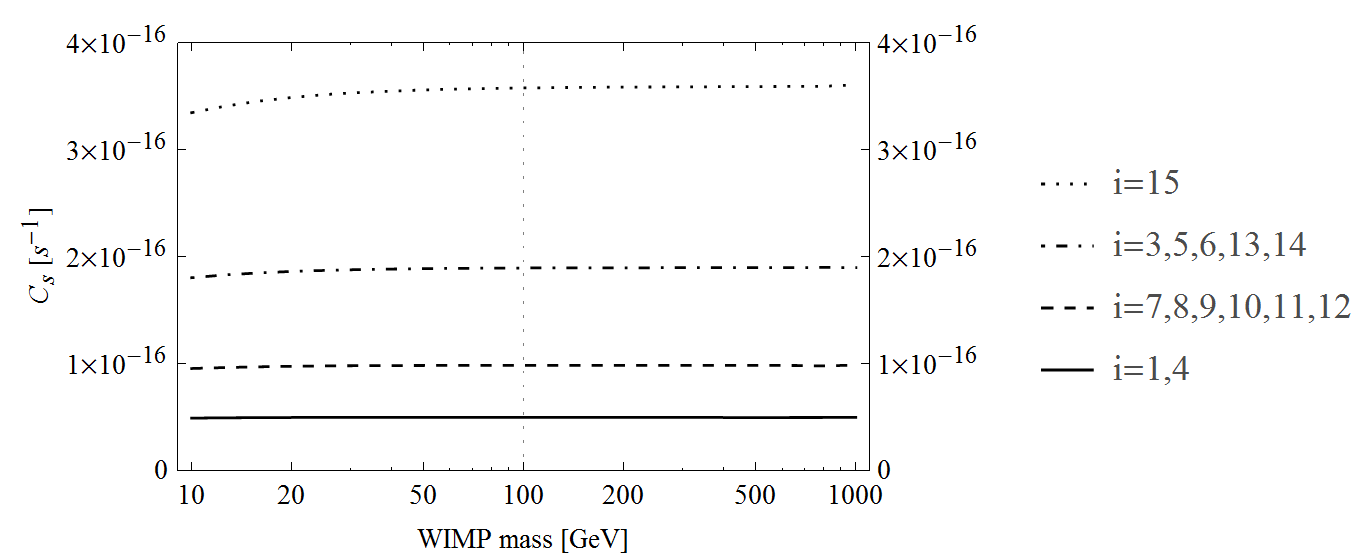}
  \end{center}
  \caption{Self-capture rates, $C_s$, for the different operators, $\hat{\mathcal{O}}_k$ in Tab.~\ref{tab:operators}.~The coupling constants are set such that they saturate the halo shapes upper limit, i.e.~$\sigma_{\chi\chi}(w=1000~\text{km}/\text{s})=1.78 \times 10^{-25} m_\chi/\text{GeV}$.~Operators represented by the same line in the figure differ in terms of capture rate $C_s$ by less than 3 \%.}
  \label{self-capture-rates}
\end{figure}

\subsection{Numerical results}
In this section we numerically compute the amplification factor $\beta(t)$ at $t=t_\odot$, where $t_\odot$ is the age of the Sun.~Results are presented in terms of curves of constant $\beta$, i.e.~iso-$\beta$ contours in the $\left(\sigma_{\chi N},\sigma_{\chi \chi}\right)$ plane, where $\sigma_{\chi N}$ and $\sigma_{\chi \chi}$ are the dark matter-nucleon scattering cross-section and dark matter self-scattering cross-section at $w=1000$~km/s, respectively.~Alternatively, we could have presented our results in the ($c_k^\tau,c_{k'}$) plane, for different $k$, $k'$ and $\tau$.~Computing our iso-$\beta$ contours, we consider different dark matter self-interaction operators.~As far as the dark matter-nucleon interaction is concerned, we focus on the isoscalar component of the $\mathcal{O}_1$ operator only (that is $c_1^1=0$).~This choice allows us to illustrate a variety of scenarios in a reasonable number of figures.~For other dark matter-nucleon interaction operators, iso-$\beta$ contours and non-equilibrium regions keep the same shape in the $\left(\sigma_{\chi N},\sigma_{\chi \chi}\right)$ plane, but are translated horizontally depending on the dark matter-nucleon operator in analysis.~The only qualitative difference between different dark matter-nucleon operators is that the excluded areas of parameter space are shifted with respect to the iso-$\beta$ lines.~Thus the non-excluded region of parameter space that is in equilibrium has different width for the various dark matter-nucleon operators.

For definiteness, we focus on spin 1/2 WIMPs with mass in the range 10--1000 GeV.~For WIMPs of spin 0 or spin 1, qualitatively similar iso-$\beta$ contours were found.~The dark matter particle spin only enters in Eq.~(\ref{eq:M2}) in Sec.~\ref{sec:theory} and in Eq.~(\ref{eq:R}) in Appendix~\ref{sec:appDM}.~In all figures presented in this section, we consider a reference WIMP thermally averaged annihilation cross-section of $\langle \sigma_{\rm ann} v_{\rm rel} \rangle = 2 \times 10^{-26}$~cm$^{3}$/s~\cite{Steigman:2012nb}.~This assumption affects the value of $\beta$ via $C_a$.~Finally, for the local population of Milky Way dark matter particles, we assume a Maxwell-Boltzmann distribution with local WIMP density, $\bar{v}$ and $v_\odot$ set as in the previous section.

In order to evaluate $\beta(t_\odot)$, we have first computed $C_c$ and $C_s$.~The total rate of dark matter capture by nuclei, $C_c$, is calculated numerically for all operators in Tab.~\ref{tab:operators}, although as already stressed at the beginning of this section, results are presented for the isoscalar component of the $\hat{\mathcal{O}}_1$ operator only.~The 16 most abundant elements in the Sun are considered in this calculation.~As already shown in~\cite{Catena:2015uha}, the element that contributes the most to the capture rate $C_c$ varies with interaction operator and, to a lesser extent, with the dark matter particle mass.~Depending on these parameters, the most important element in the capture is H, $^4$He, $^{14}$N, $^{16}$O, $^{27}$Al, $^{56}$Fe, or $^{59}$Ni.~For some interaction operators the heavier nuclei are the most important ones, especially for operators with a quadratic or cubic dependence on momentum transfer.

\begin{figure}
\begin{center}
\begin{minipage}[t]{0.49\linewidth}
\centering
\includegraphics[width=\textwidth]{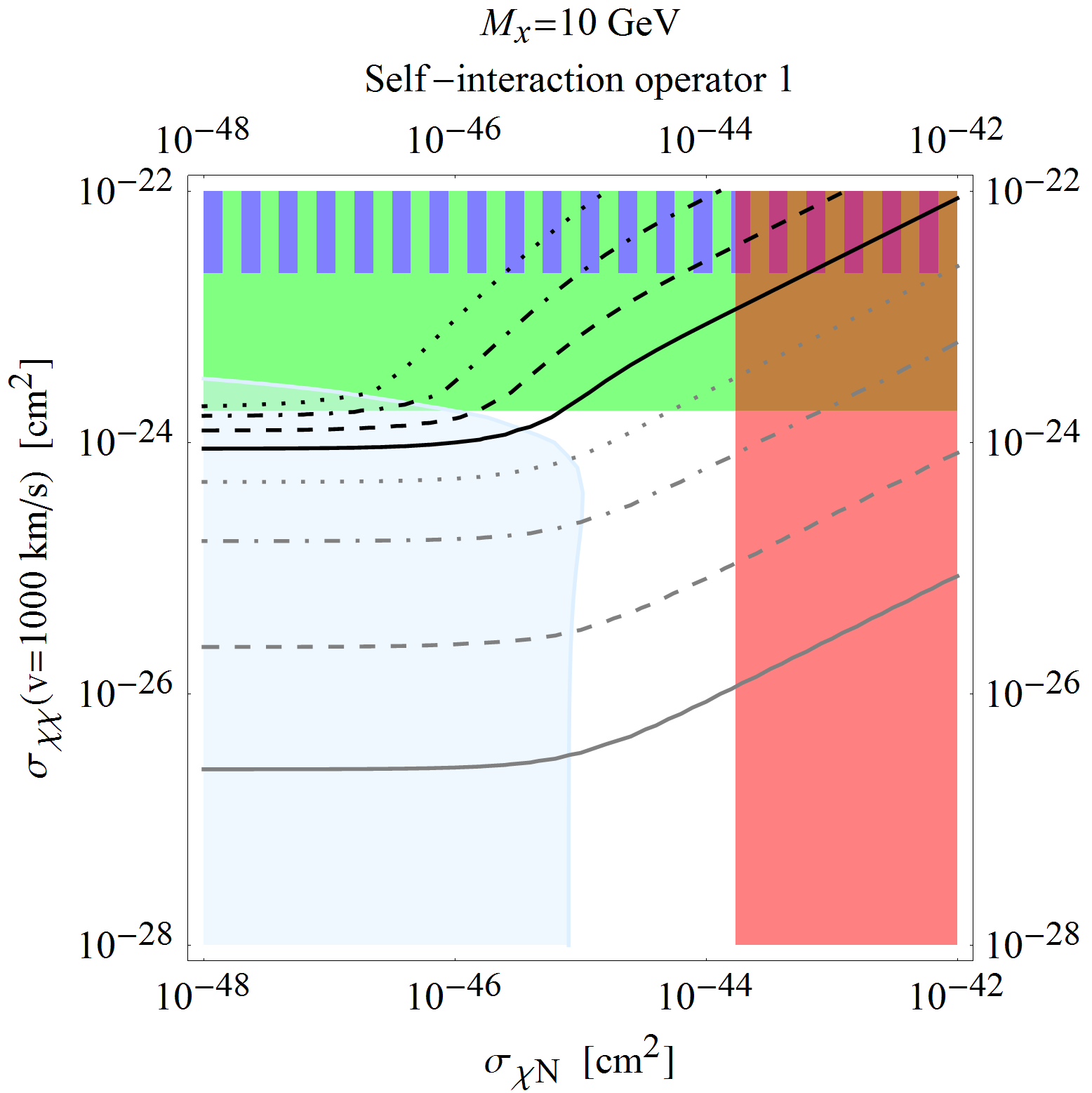}
\end{minipage}
\begin{minipage}[t]{0.49\linewidth}
\centering
\includegraphics[width=\textwidth]{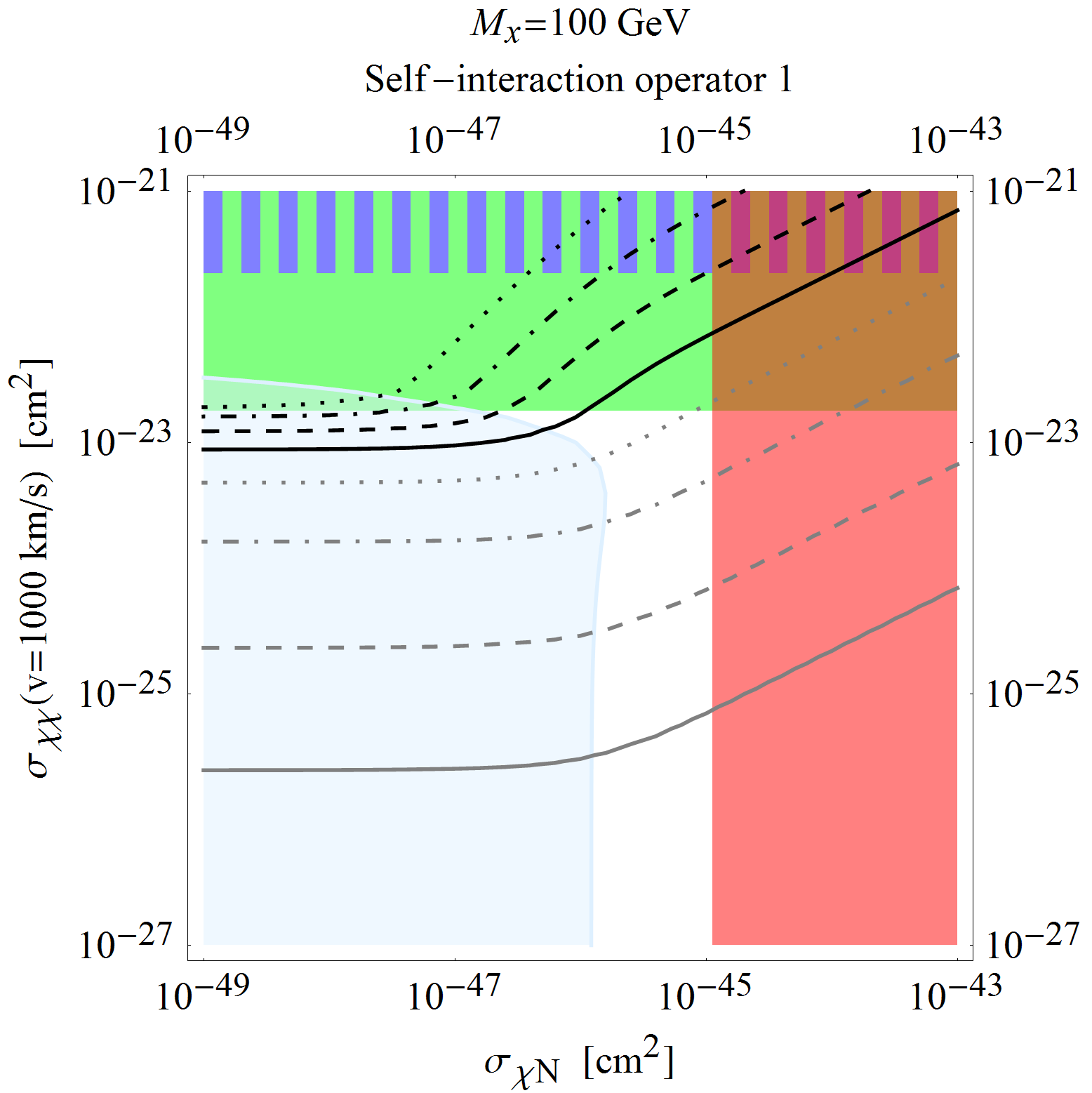}
\end{minipage}
\begin{minipage}[t]{0.49\linewidth}
\centering
\vspace*{0.05cm}\includegraphics[width=\textwidth]{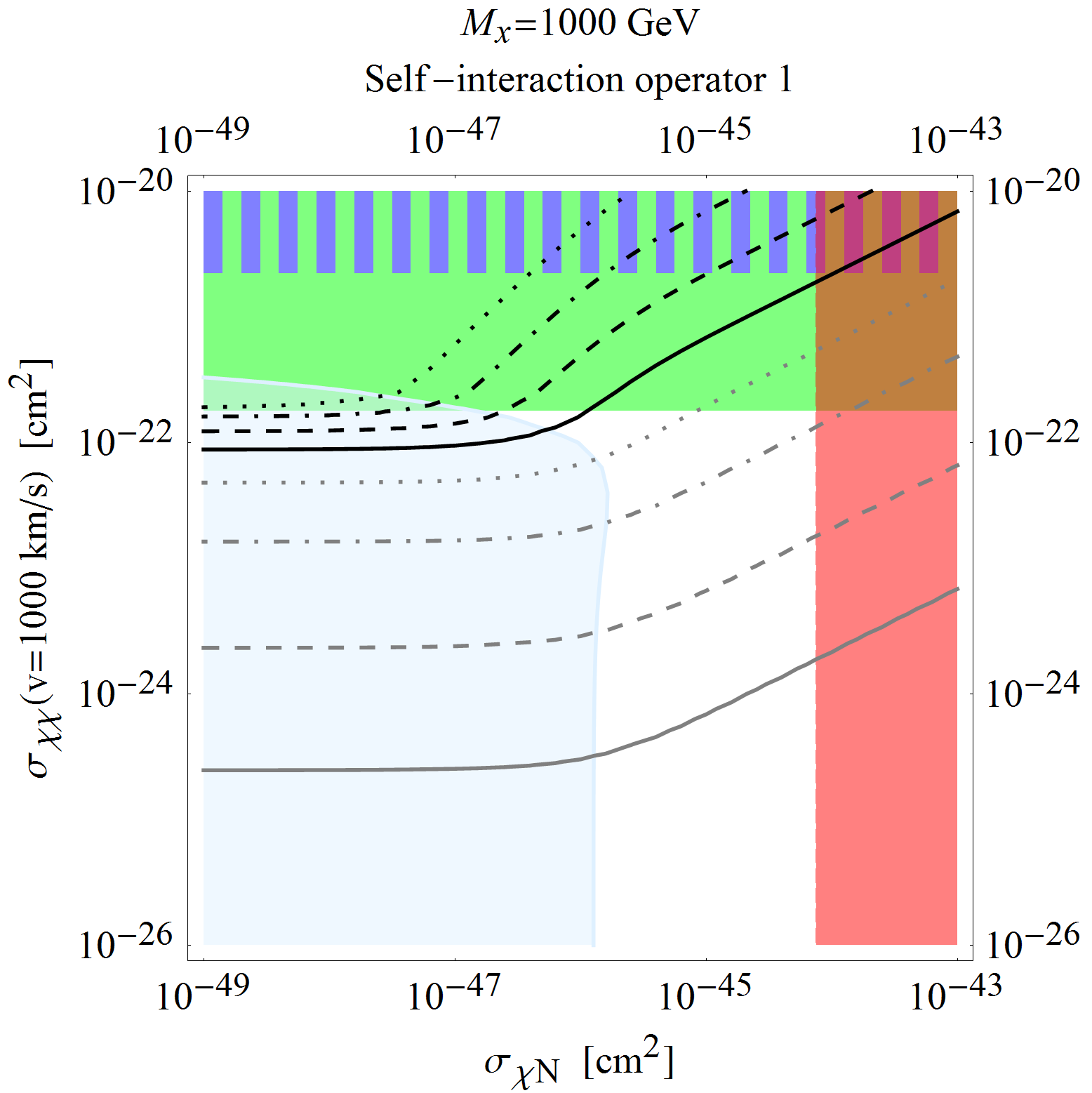}
\end{minipage}
\begin{minipage}[t]{0.49\linewidth}
\centering
\vspace*{2cm}\includegraphics[width=0.65\textwidth]{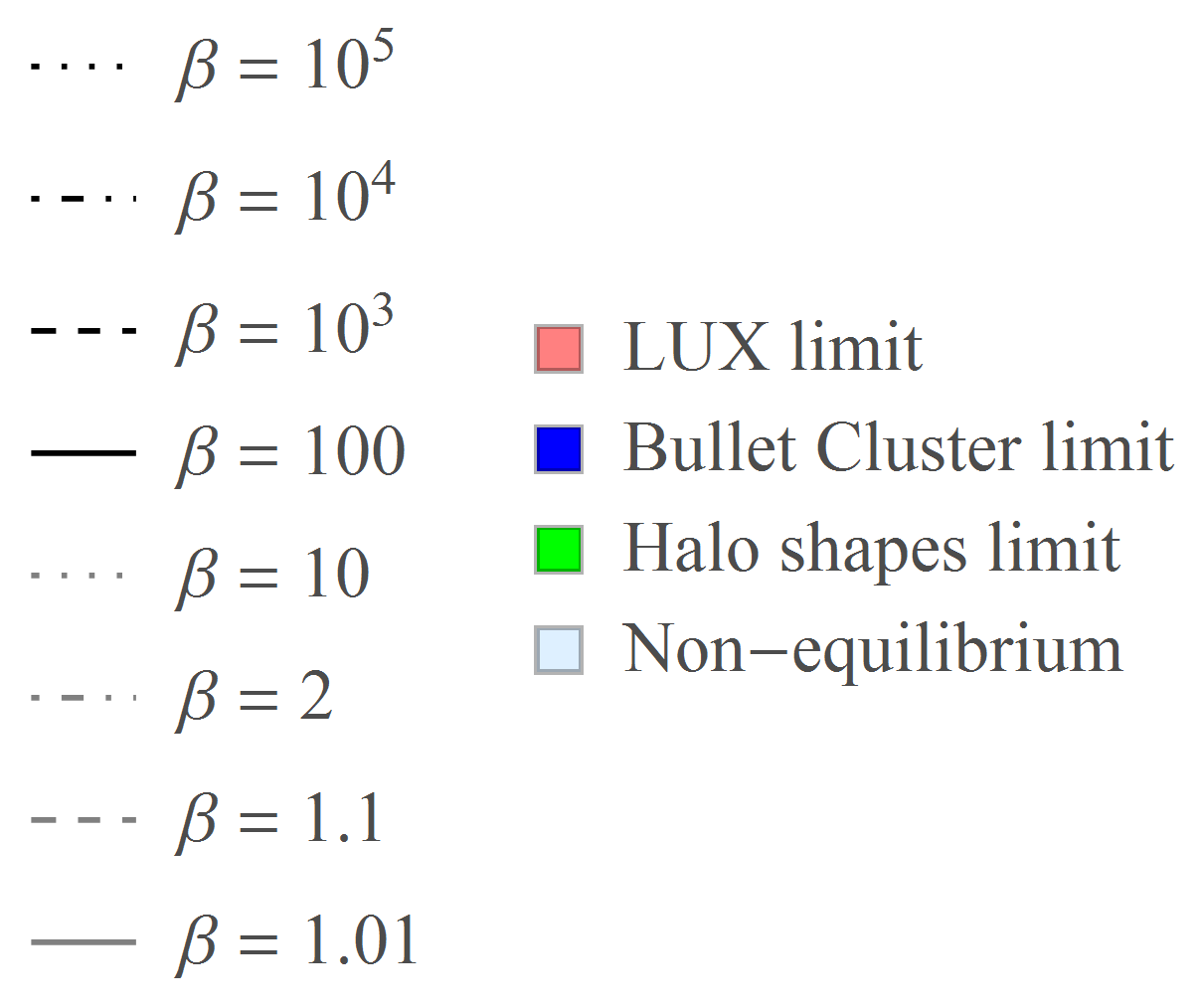}
\end{minipage}
\end{center}
\caption{Lines of constant amplification, $\beta$, in the ($\sigma_{\chi N}, \sigma_{\chi \chi}$) plane.~We denote by $\sigma_{\chi N}$ the WIMP-nucleon scattering cross-section for the interaction operator $\hat{\mathcal{O}}_1$ ($c_1^1=0$), and by $\sigma_{\chi\chi}$ the WIMP self-interaction cross-section for the self-interaction operator $\hat{\mathcal{O}}_1$.~Different panels refer to distinct WIMP masses, as shown in the legends.~Coloured regions represent experimental exclusion limits and the non-equilibrium area.}
\label{betalines-operator-1}
\end{figure}

\begin{figure}
\begin{center}
\begin{minipage}[t]{0.49\linewidth}
\centering
\includegraphics[width=\textwidth]{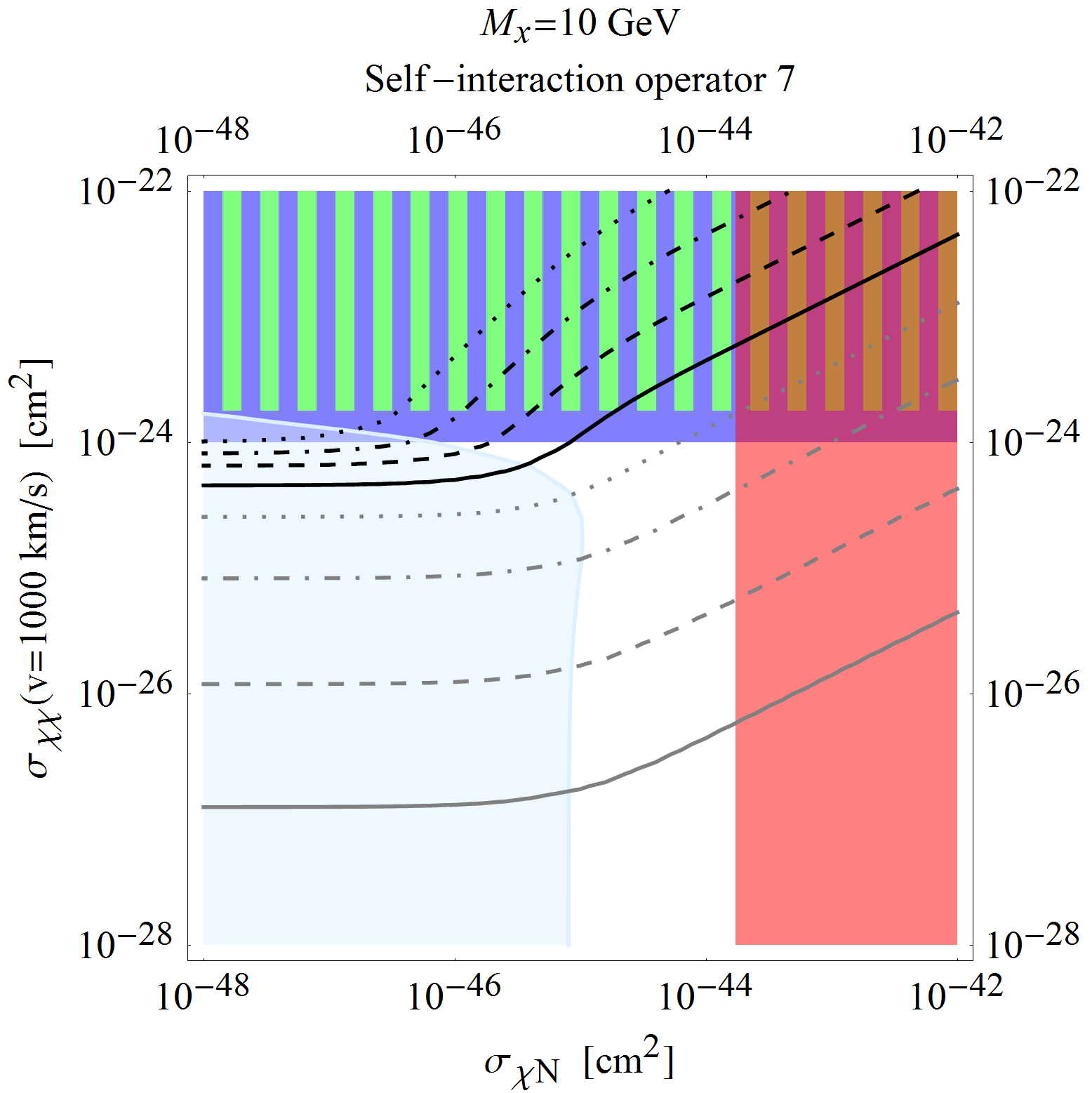}
\end{minipage}
\begin{minipage}[t]{0.49\linewidth}
\centering
\includegraphics[width=\textwidth]{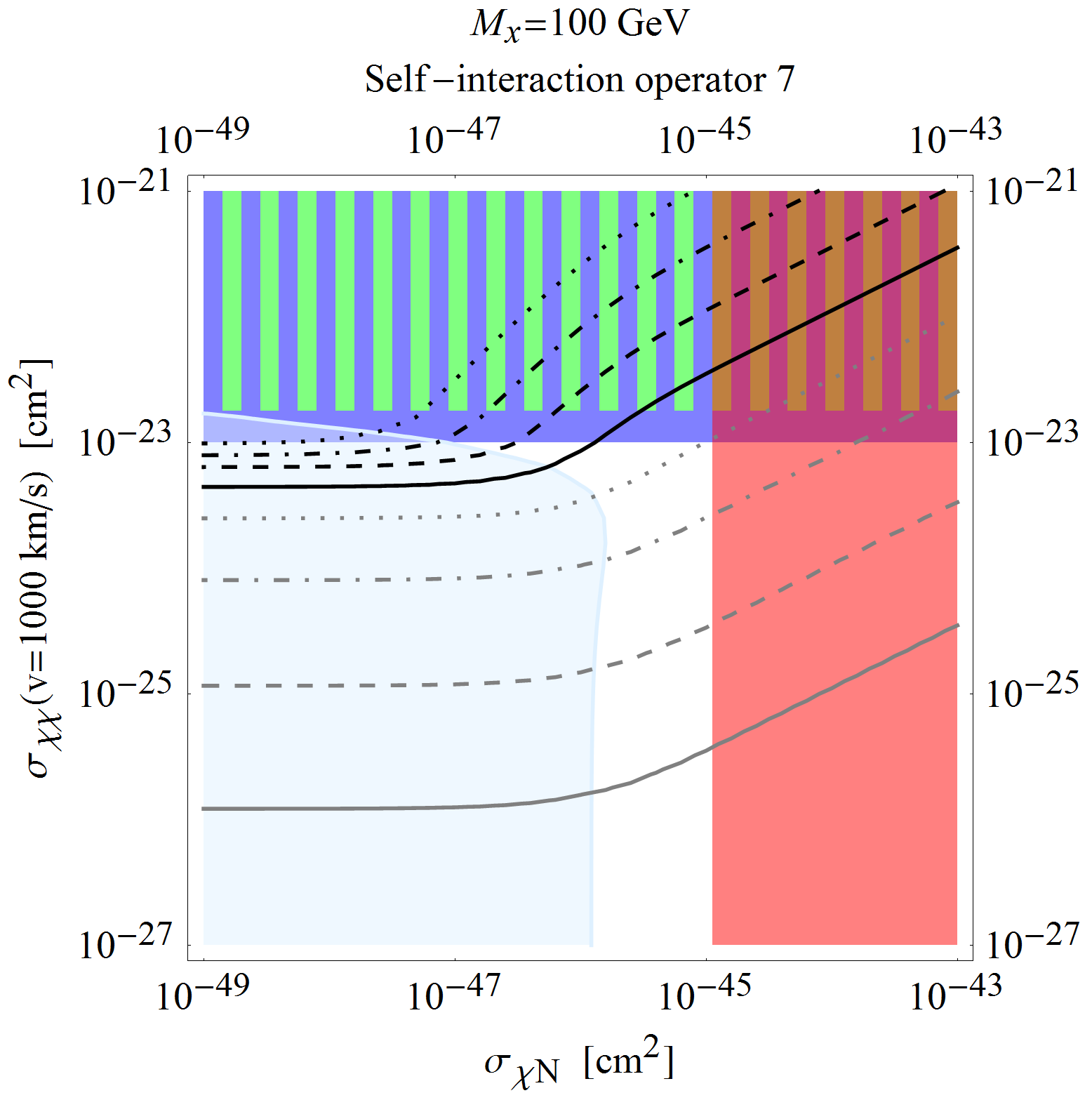}
\end{minipage}
\begin{minipage}[t]{0.49\linewidth}
\centering
\vspace*{0.05cm}\includegraphics[width=\textwidth]{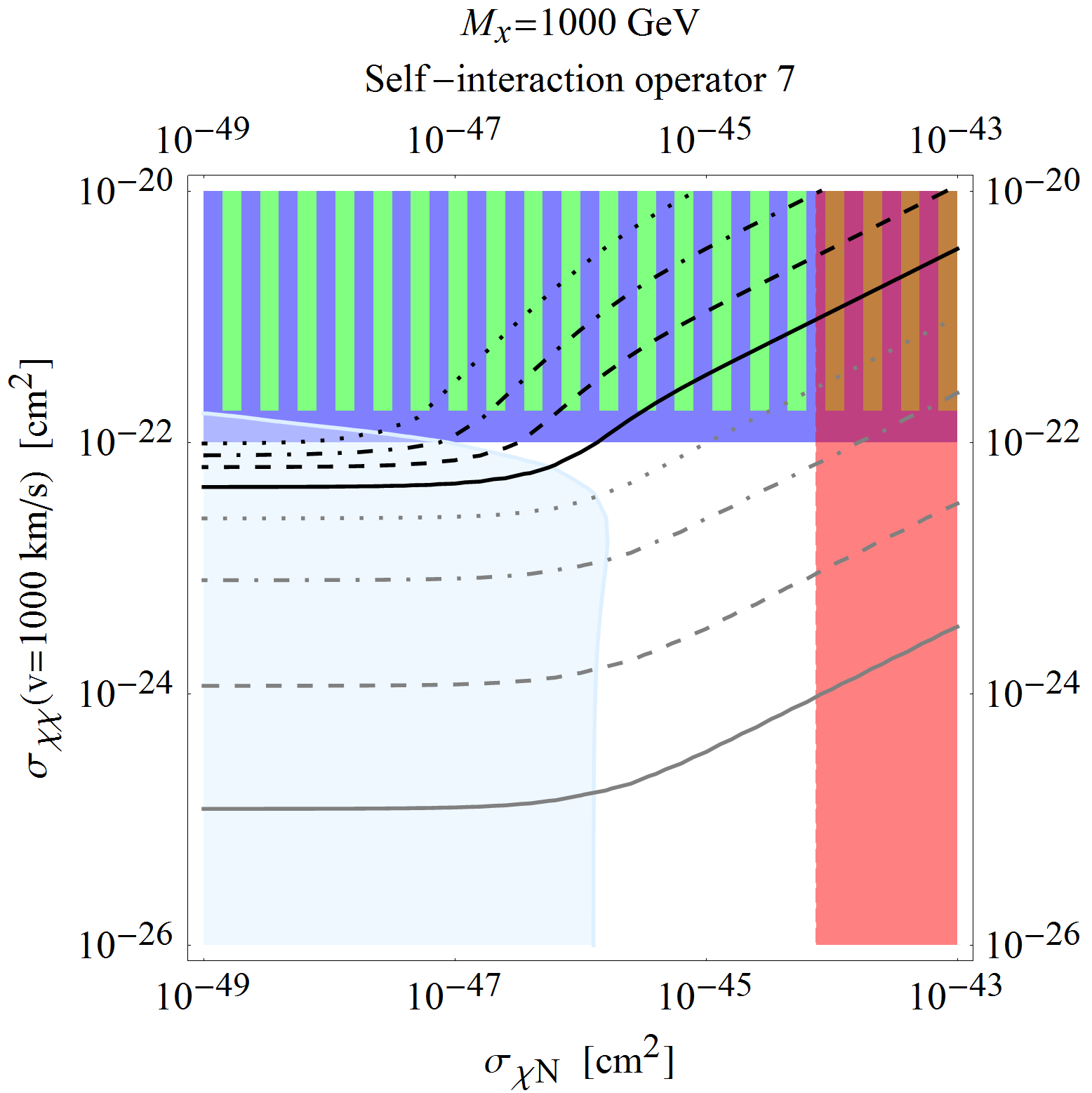}
\end{minipage}
\begin{minipage}[t]{0.49\linewidth}
\centering
\vspace*{2cm}\includegraphics[width=0.65\textwidth]{./Legend.png}
\end{minipage}
\end{center}
\caption{Same as for Fig.~\ref{betalines-operator-1}, but now for the dark matter-nucleon interaction operator $\hat{\mathcal{O}}_1$ and the dark matter self-interaction operator $\hat{\mathcal{O}}_7$.}
\label{betalines-operator-7}
\end{figure}

\begin{figure}
\begin{center}
\begin{minipage}[t]{0.49\linewidth}
\centering
\includegraphics[width=\textwidth]{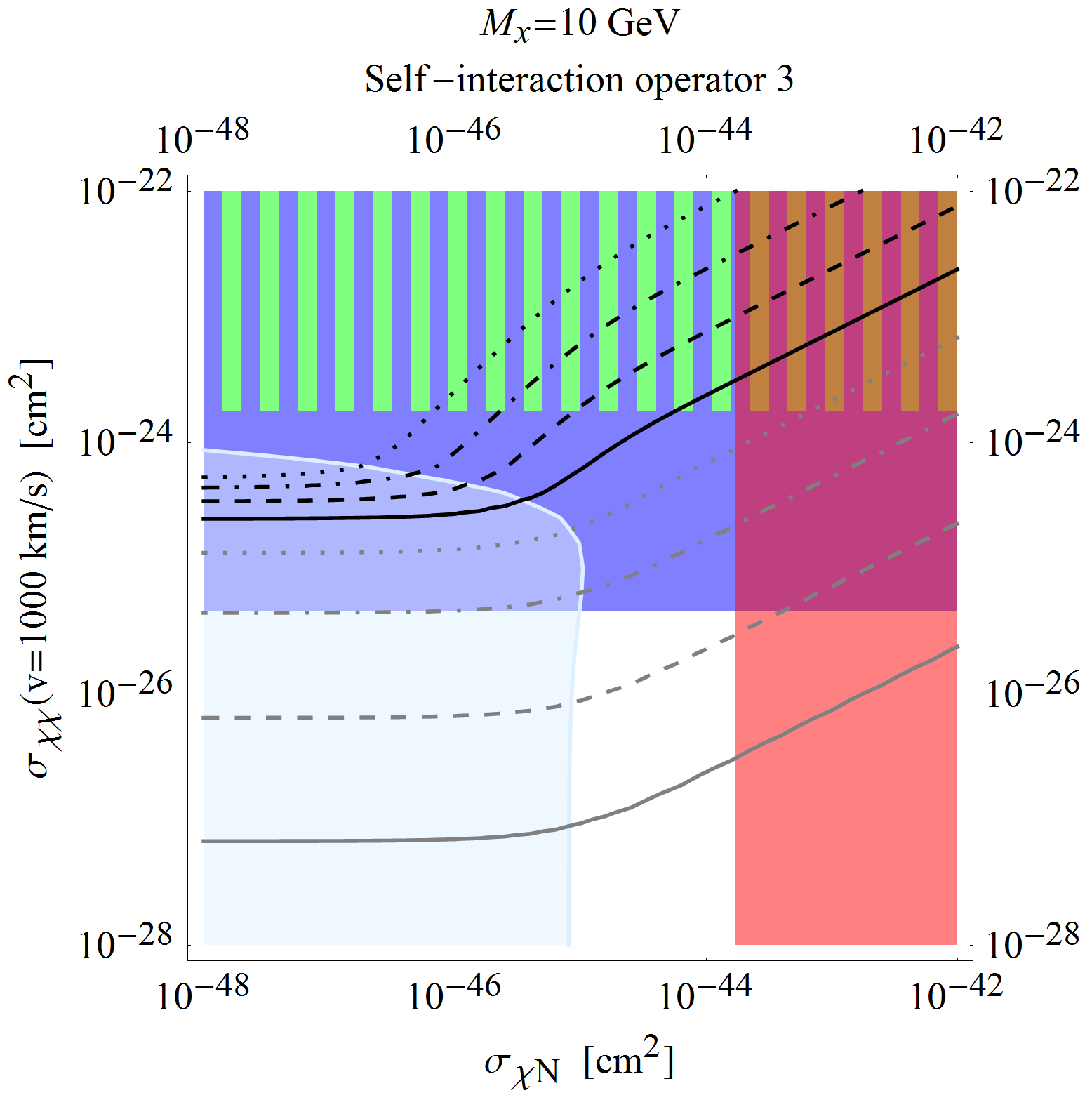}
\end{minipage}
\begin{minipage}[t]{0.49\linewidth}
\centering
\includegraphics[width=\textwidth]{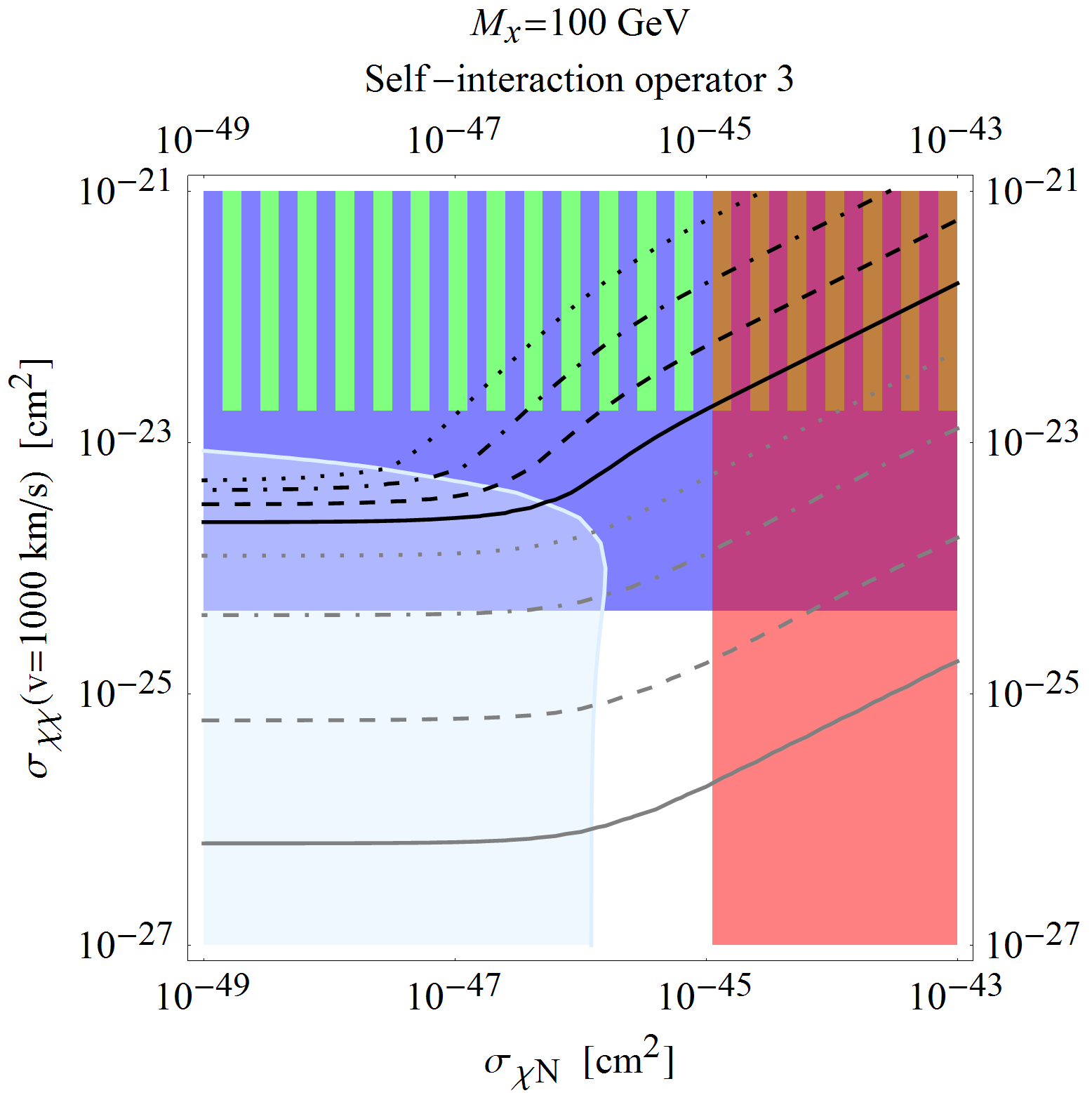}
\end{minipage}
\begin{minipage}[t]{0.49\linewidth}
\centering
\vspace*{0.05cm}\includegraphics[width=\textwidth]{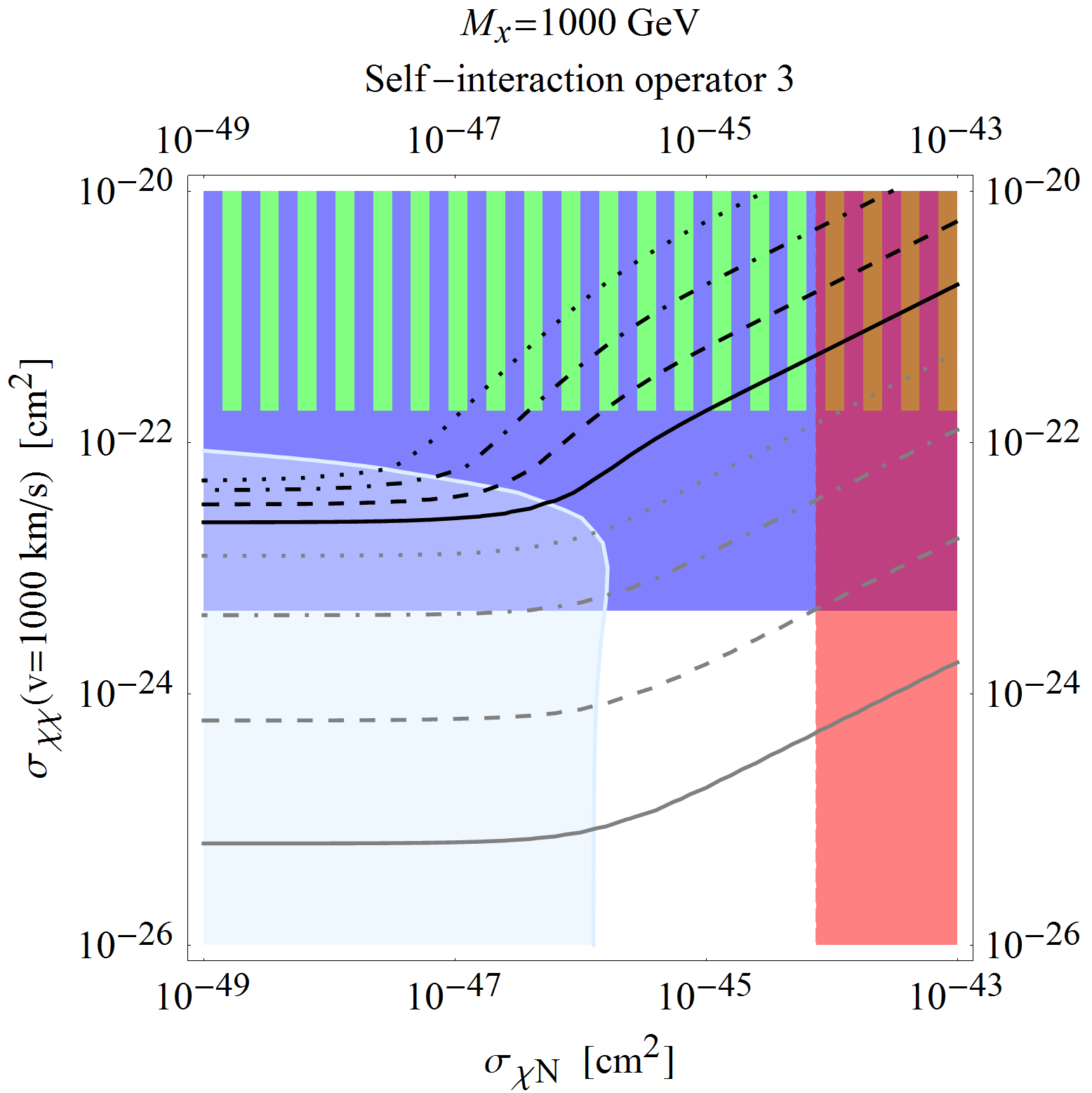}
\end{minipage}
\begin{minipage}[t]{0.49\linewidth}
\centering
\vspace*{2cm}\includegraphics[width=0.65\textwidth]{./Legend.png}
\end{minipage}
\end{center}
\caption{Same as for Fig.~\ref{betalines-operator-1}, but now for the dark matter-nucleon interaction operator $\hat{\mathcal{O}}_1$ and the dark matter self-interaction operator $\hat{\mathcal{O}}_3$.}
\label{betalines-operator-3}
\end{figure}

\begin{figure}
\begin{center}
\begin{minipage}[t]{0.49\linewidth}
\centering
\includegraphics[width=\textwidth]{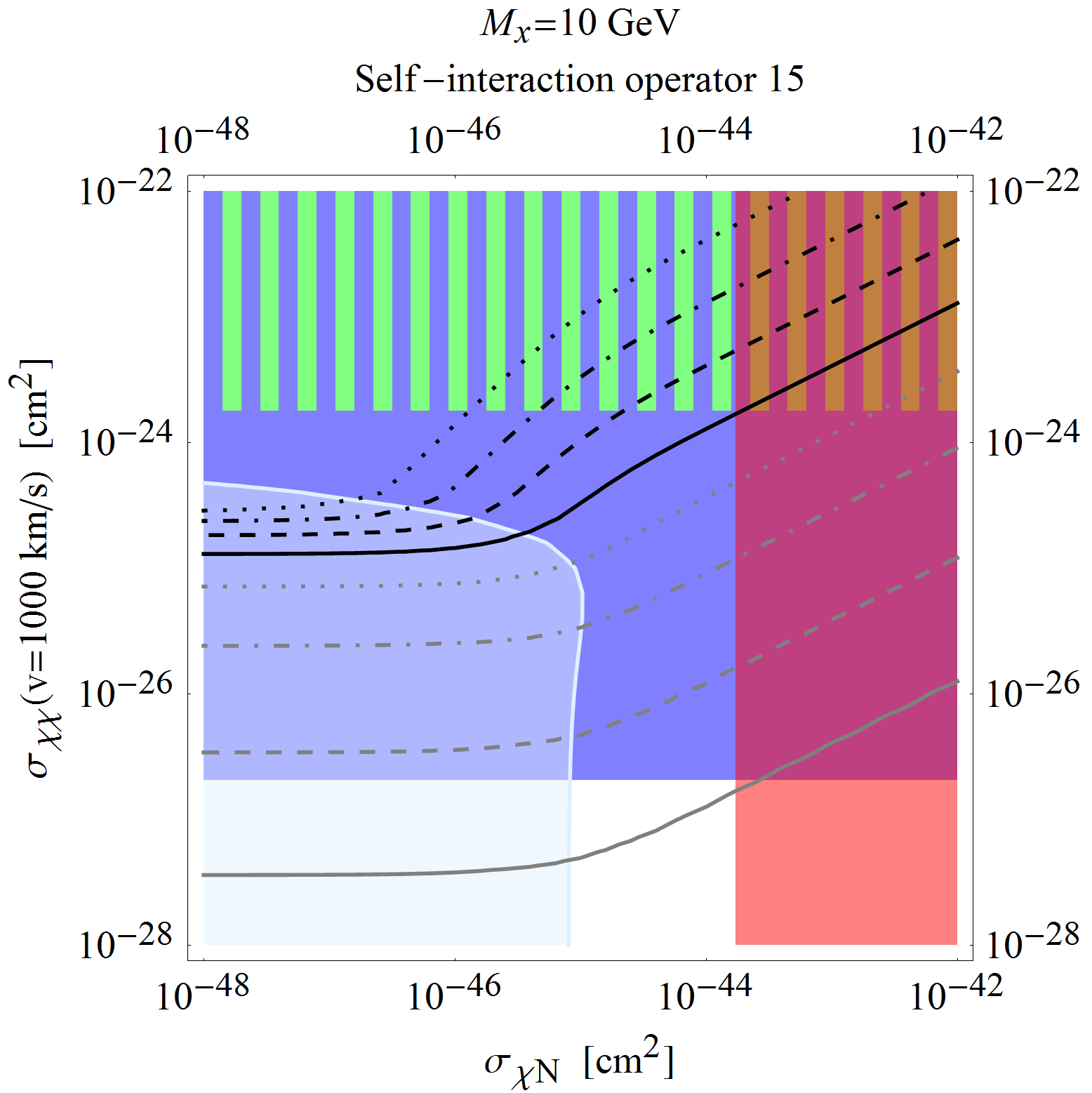}
\end{minipage}
\begin{minipage}[t]{0.49\linewidth}
\centering
\includegraphics[width=\textwidth]{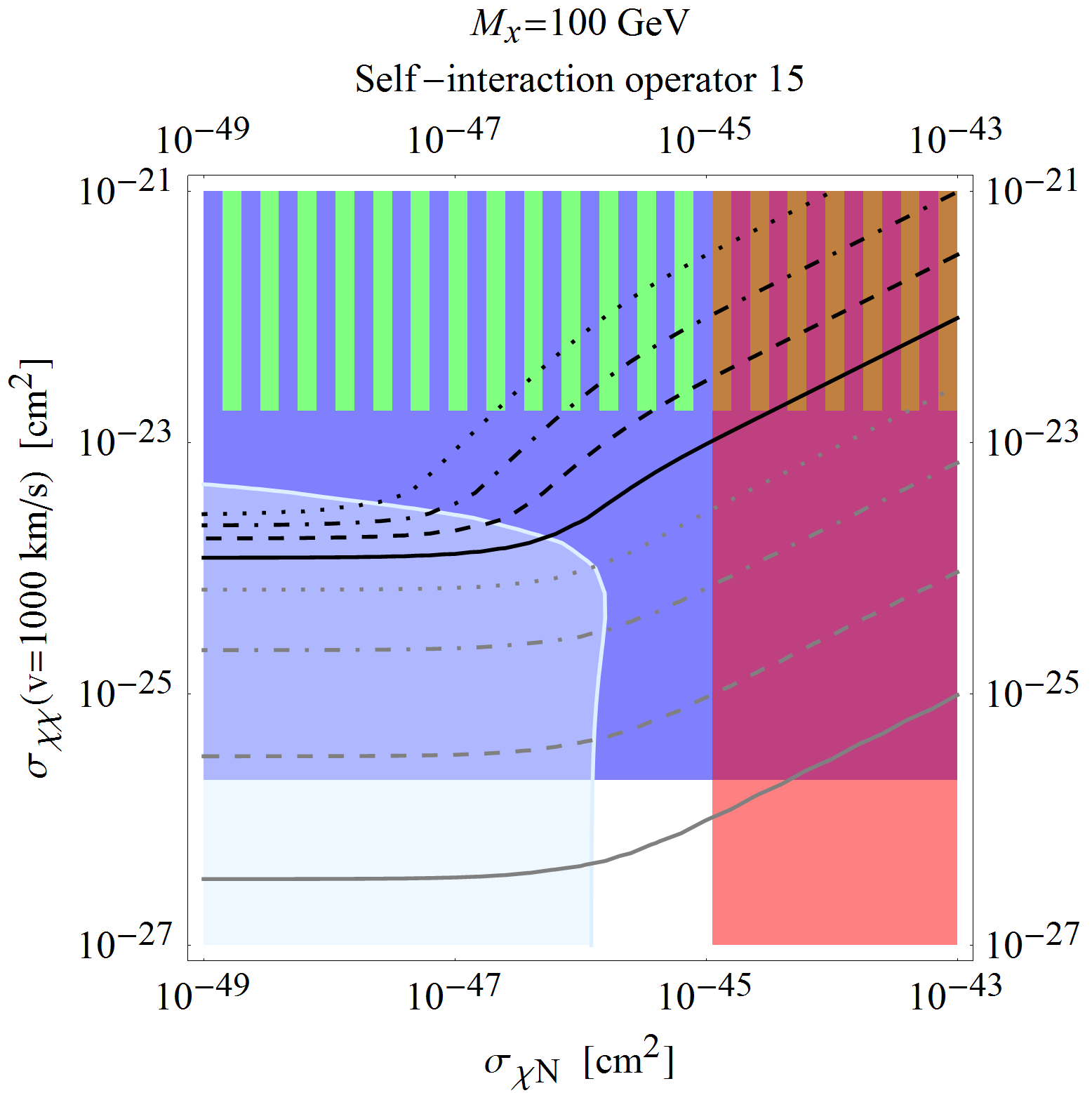}
\end{minipage}
\begin{minipage}[t]{0.49\linewidth}
\centering
\vspace*{0.05cm}\includegraphics[width=\textwidth]{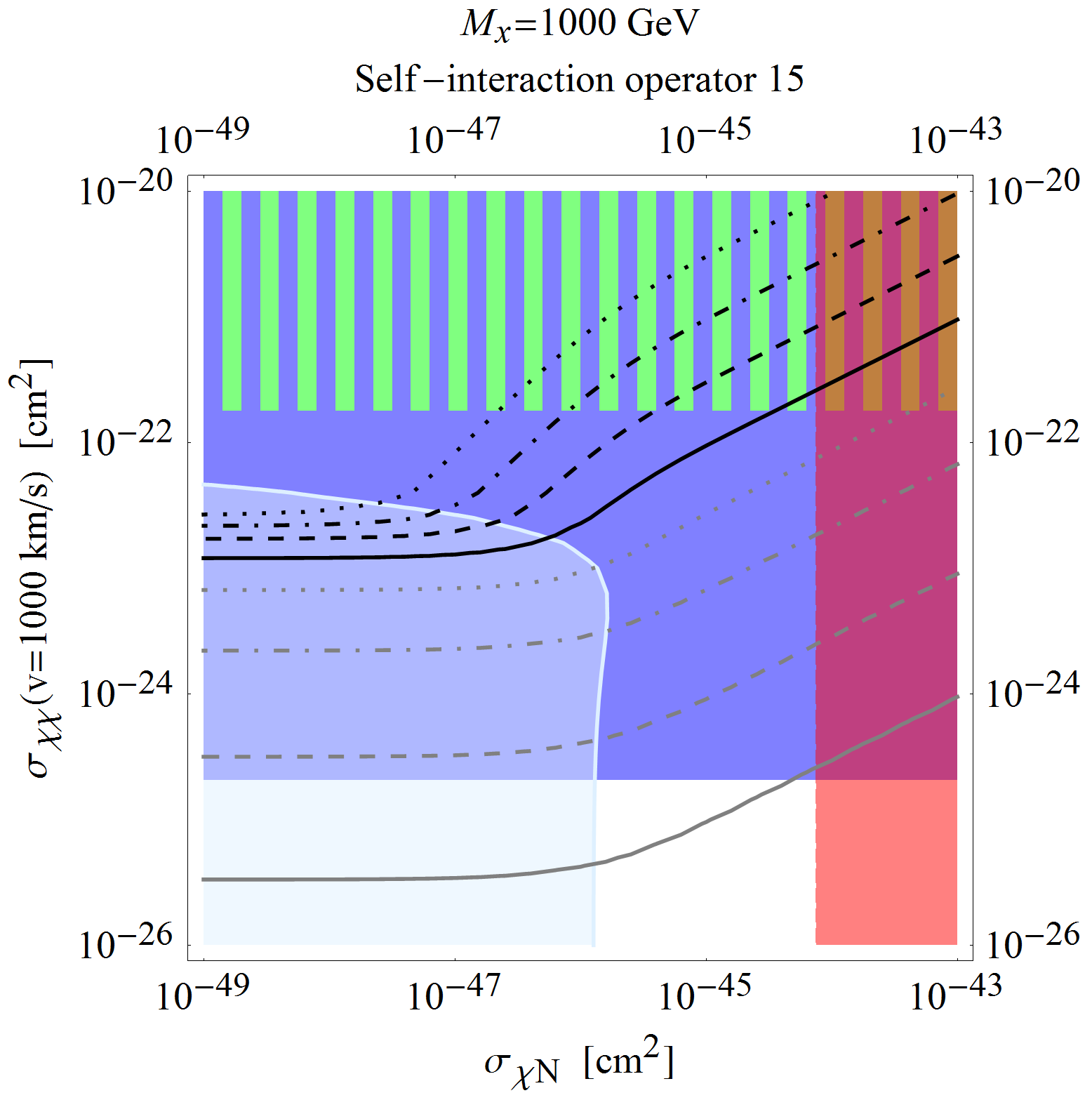}
\end{minipage}
\begin{minipage}[t]{0.49\linewidth}
\centering
\vspace*{2cm}\includegraphics[width=0.65\textwidth]{./Legend.png}
\end{minipage}
\end{center}
\caption{Same as for Fig.~\ref{betalines-operator-1}, but now for the dark matter-nucleon interaction operator $\hat{\mathcal{O}}_1$ and the dark matter self-interaction operator $\hat{\mathcal{O}}_{15}$.}
\label{betalines-operator-15}
\end{figure}
 
The total rate of dark matter capture by the Sun via self-interaction, $C_s$, is calculated for the 14 operators in Tab.~\ref{tab:operators}.~Fig.~\ref{self-capture-rates} shows $C_s$ as a function of $m_\chi$ for the different self-interaction operators.~For each operator, the corresponding coupling constant $c_k$ is set at its maximum value given by $\sigma_{\chi\chi}(w=1000~\text{km}/\text{s})=1.78 \times 10^{-25}$ $m_\chi/\text{GeV}\;\text{cm}^2$.~Interactions fall into four distinct groups, depending on the power of ${\bf {\hat{q}}}$ in the corresponding operator, and therefore depending on how $\sigma_{\chi\chi}$ varies with $w$:~$\sigma_{\chi\chi}\propto w^0$ for $\hat{\mathcal{O}}_1$ and $\hat{\mathcal{O}}_4$; $\sigma_{\chi\chi}\propto w^2$ for $\hat{\mathcal{O}}_7, \dots, \hat{\mathcal{O}}_{12}$; $\sigma_{\chi\chi}\propto w^4$ for $\hat{\mathcal{O}}_3$, $\hat{\mathcal{O}}_{5}$, $\hat{\mathcal{O}}_6$, $\hat{\mathcal{O}}_{13}$ and $\hat{\mathcal{O}}_{14}$; and $\sigma_{\chi\chi}\propto w^6$ for $\hat{\mathcal{O}}_{15}$.~Notice that the velocity integral in the definition of $C_s$ peaks at $u\rightarrow 0$ which implies $w \sim v(0)\simeq 1380$~km/s in Eq.~(\ref{selfcapturesigma}).~As a result, the self-scattering cross-section normalisation $\sigma_{\chi\chi}(w=1000~\text{km}/\text{s})=1.78 \times 10^{-25}$~$m_\chi/\text{GeV}\;\text{cm}^2$, with $w=1000~{\rm km}/{\rm s}<v(0)$, implies the hierarchy of capture rates in Fig.~\ref{self-capture-rates}.

An inverted hierarchy of capture rates arises when self-scattering cross-sections are normalised to the Bullet cluster limit in Eq.~(\ref{eq:bullet}), since in this case $v(0)<w=4700~{\rm km}/{\rm s}$, which is the collisional velocity of the Bullet cluster.~Overall, the Bullet cluster limit,  Eq.~(\ref{eq:bullet}), and the limit on the self-scattering cross-section from halo shapes, Eq.~(\ref{v=1000limit}), are complementary.~For interaction operators with no dependence on the momentum transfer ($\hat{\mathcal{O}}_1$ and $\hat{\mathcal{O}}_4$), the strongest limit comes from N-body simulations (where $w=1000$ km/s); for operators with quadratic or cubic dependence on the momentum transfer, the strongest limit comes from the Bullet cluster (where $w=4700$~km/s); for interaction operators with a linear dependence on the momentum transfer, the two limits are about equal.

Figs.~\ref{betalines-operator-1}, \ref{betalines-operator-7}, \ref{betalines-operator-3} and \ref{betalines-operator-15} show the iso-$\beta$ contours that we obtain in the $\left(\sigma_{\chi N},\sigma_{\chi \chi}\right)$ plane for dark matter self-interaction operators of type $\hat{\mathcal{O}}_1$, $\hat{\mathcal{O}}_7$, $\hat{\mathcal{O}}_3$ and $\hat{\mathcal{O}}_{15}$, respectively.~The four self-interaction operators have been chosen in order to represent the four behaviours illustrated in Fig.~\ref{self-capture-rates}.~In all panels, we consider the dark matter-nucleon interaction $\hat{\mathcal{O}}_1$ with $c_1^0\neq0$ and $c_1^1=0$.~The amplification factor is calculated at present time, $\beta(t=t_\odot)$, which implies that for large portions of the $\left(\sigma_{\chi N},\sigma_{\chi \chi}\right)$ plane equilibrium has not yet been reached.~The self-scattering cross-section reported in the panels is evaluated at the reference WIMP velocity of 1000~km/s.

In Figs.~\ref{betalines-operator-1}, \ref{betalines-operator-7}, \ref{betalines-operator-3} and \ref{betalines-operator-15}, red regions are excluded by limits on the dark matter-nucleon scattering cross-section derived in~\cite{Catena:2015iea} from LUX data.~Blue and green regions are excluded by limits on the dark matter self-interaction strength from an analysis of the Bullet cluster~\cite{BulletClusterLimit08} and halo shape analyses in N-body simulations~\cite{Rocha:2012jg}, respectively.~The light blue region represents the area of non-equilibrium.~It is the region in parameter space where the equilibrium solution to Eq.~(\ref{diffeqwithself}), namely, Eq.~(\ref{betaequilibriumsolution}), is not reached within a time interval of $t_\odot\simeq4.5\times10^9$ years, corresponding to the age of the Sun.~Following~\cite{Zentner:2009is}, we approximate the non-equilibrium region as the contour in parameter space corresponding to a neutrino flux less than 58\% of the corresponding flux at equilibrium.~The specific value 58\% is related to the fact that, for $C_s=0$, the non-equilibrium to equilibrium neutrino flux ratio is equal to $\tanh^2 (t\sqrt{C_c C_a})$, and for $t_{\rm eq}\equiv 1/\sqrt{C_c C_a}$, such ratio equals $\tanh^2(1)\simeq 0.58$.~The iso-$\beta$ contours in the figures are very different in the equilibrium and non-equilibrium regions.~In the former, they follow the approximate relation $\sigma_{\chi N} \propto \sigma_{\chi\chi}^2$, as expected from Eq.~(\ref{betaequilibriumsolution}).~In the latter, the iso-$\beta$ contours are independent of the dark matter-nucleon scattering cross-section, $\sigma_{\chi N}$, in agreement with Eq.~(\ref{betanonequilibriumsolution}).

For operators with no dependence on the momentum transfer, i.e.~$\hat{\mathcal{O}}_1$ and $\hat{\mathcal{O}}_4$, a signal amplification of several orders of magnitude is allowed, as shown in Fig.~\ref{betalines-operator-1}.~This is also true for operators with a linear dependence on the momentum transfer (those labelled by $k=7,8,9,10,11,12$), as illustrated in Fig.~\ref{betalines-operator-7}.~In these cases it could be possible that self-interaction completely dominates the total rate of dark matter capture by the Sun.~For operators with a quadratic dependence on the momentum transfer (those labelled by $k=3,5,6,13,14$), the Bullet cluster limit allows a smaller signal amplification, as apparent in Fig.~\ref{betalines-operator-3}.~In this case, dark matter capture via scattering by nuclei would be the dominant process.~Finally, for the operator $\hat{\mathcal{O}}_{15}$, which is characterised by an indirect cubic dependence on the momentum transfer through the operator ${\bf{\hat{v}}}^{\perp}$, the Bullet cluster limit excludes any signal amplification larger than a few percent, as shown in Fig.~\ref{betalines-operator-15}.~In this case, dark matter self-interaction has a negligible effect on the total capture rate.

\section{Conclusions}
\label{sec:conclusions}
We studied the capture of WIMP dark matter by the Sun in non-relativistic effective theories for dark matter-nucleon and dark matter self-interactions mediated by a heavy particle.~Whereas effective theories were already used previously to model the scattering of WIMPs by nuclei in the Sun, this is the first time that an effective theory approach to dark matter self-interactions is used in this context.~This general theoretical framework allowed us to perform a model independent analysis of the capture of dark matter particles by the Sun via self-interaction.

Within this theoretical framework, we compared the total rate of dark matter capture by the Sun in presence of dark matter self-interactions to the same quantity when dark matter self-scattering in the solar interior is neglected.~The time dependent ratio of these two rates, denoted by $\beta$ in Eq.~(\ref{beta}), determines the relative enhancement of the neutrino flux from WIMP annihilation in the Sun due to dark matter self-interaction.~We computed this ratio at present time for all dark matter-nucleon and dark matter self-interaction operators in Tab.~\ref{tab:operators}.~Results were presented for selected dark matter-nucleon and self-interaction operators in the $\left(\sigma_{\chi N},\sigma_{\chi \chi}\right)$ plane, where $\sigma_{\chi N}$ and $\sigma_{\chi \chi}$ are the dark matter-nucleon scattering cross-section and dark matter self-scattering cross-section, respectively.~Limits from the LUX experiment, the Bullet cluster and halo shape analyses in N-body simulations were imposed on this parameter space.
 
We found that for self-interaction operators with no dependence on the momentum transfer $\hat{\bf{q}}$, that is, $\hat{\mathcal{O}}_1$ and $\hat{\mathcal{O}}_4$, a signal amplification of several orders of magnitude is possible, even for standard thermally averaged annihilation cross-sections $\langle \sigma_{\rm ann} v_{\rm rel}\rangle$.~This is in contrast with previous findings~\cite{Zentner:2009is}, which at the same time do not seem to numerically reproduce the analytic dependence on the capture rate $C_c$ given in Eq.~(\ref{betaequilibriumsolution}) of this work, and in Eq.~(7) of~\cite{Zentner:2009is}.~This significant amplification is reported in Fig.~\ref{betalines-operator-1}.~We also found a large amplification for self-interaction operators with a linear dependence on the momentum transfer (those labelled by $k=7,8,9,10,11,12$).~Results for these operators are given in Fig.~\ref{betalines-operator-7}.~For the operators in Figs.~\ref{betalines-operator-1} and \ref{betalines-operator-7}, self-interactions could completely dominate the rate of WIMP capture by the Sun.~For operators with a quadratic dependence on the momentum transfer (those labelled by $k=3,5,6,13,14$), the Bullet cluster limit allows for a smaller signal amplification, whereas for the operator $\hat{\mathcal{O}}_{15}$, that exhibits an indirect cubic dependence on the momentum transfer through the operator ${\bf{\hat{v}}}^{\perp}$, the Bullet cluster limit excludes any signal amplification larger than a few percent.~The latter two cases are discussed in Fig.~\ref{betalines-operator-3} and Fig.~\ref{betalines-operator-15}, respectively.

The results found in this investigation can be used to optimise present and future searches for dark matter self-interactions at neutrino telescopes.~At the same time, our findings lay the foundations for model independent studies of the dark matter capture by the Sun via self-interaction.

\appendix

\section{Dark matter response functions}
\label{sec:appDM}
Using the same notation adopted in the previous sections, below we list the dark matter response functions that appear in Eq.~(\ref{eq:sigma}):
\begin{eqnarray}
 R_{M}^{\tau \tau^\prime}\left(v_T^{\perp 2}, {q^2 \over m_N^2}\right) &=& c_1^\tau c_1^{\tau^\prime } + {J_\chi (J_\chi+1) \over 3} \left[ {q^2 \over m_N^2} v_T^{\perp 2} c_5^\tau c_5^{\tau^\prime }+v_T^{\perp 2}c_8^\tau c_8^{\tau^\prime }
+ {q^2 \over m_N^2} c_{11}^\tau c_{11}^{\tau^\prime } \right] \nonumber \\
 R_{\Phi^{\prime \prime}}^{\tau \tau^\prime}\left(v_T^{\perp 2}, {q^2 \over m_N^2}\right) &=& {q^2 \over 4 m_N^2} c_3^\tau c_3^{\tau^\prime } + {J_\chi (J_\chi+1) \over 12} \left( c_{12}^\tau-{q^2 \over m_N^2} c_{15}^\tau\right) \left( c_{12}^{\tau^\prime }-{q^2 \over m_N^2}c_{15}^{\tau^\prime} \right)  \nonumber \\
 R_{\Phi^{\prime \prime} M}^{\tau \tau^\prime}\left(v_T^{\perp 2}, {q^2 \over m_N^2}\right) &=&  c_3^\tau c_1^{\tau^\prime } + {J_\chi (J_\chi+1) \over 3} \left( c_{12}^\tau -{q^2 \over m_N^2} c_{15}^\tau \right) c_{11}^{\tau^\prime } \nonumber \\
  R_{\tilde{\Phi}^\prime}^{\tau \tau^\prime}\left(v_T^{\perp 2}, {q^2 \over m_N^2}\right) &=&{J_\chi (J_\chi+1) \over 12} \left[ c_{12}^\tau c_{12}^{\tau^\prime }+{q^2 \over m_N^2}  c_{13}^\tau c_{13}^{\tau^\prime}  \right] \nonumber \\
   R_{\Sigma^{\prime \prime}}^{\tau \tau^\prime}\left(v_T^{\perp 2}, {q^2 \over m_N^2}\right)  &=&{q^2 \over 4 m_N^2} c_{10}^\tau  c_{10}^{\tau^\prime } +
  {J_\chi (J_\chi+1) \over 12} \left[ c_4^\tau c_4^{\tau^\prime} + \right.  \nonumber \\
 && \left. {q^2 \over m_N^2} ( c_4^\tau c_6^{\tau^\prime }+c_6^\tau c_4^{\tau^\prime })+
 {q^4 \over m_N^4} c_{6}^\tau c_{6}^{\tau^\prime } +v_T^{\perp 2} c_{12}^\tau c_{12}^{\tau^\prime }+{q^2 \over m_N^2} v_T^{\perp 2} c_{13}^\tau c_{13}^{\tau^\prime } \right] \nonumber \\
    R_{\Sigma^\prime}^{\tau \tau^\prime}\left(v_T^{\perp 2}, {q^2 \over m_N^2}\right)  &=&{1 \over 8} \left[ {q^2 \over  m_N^2}  v_T^{\perp 2} c_{3}^\tau  c_{3}^{\tau^\prime } + v_T^{\perp 2}  c_{7}^\tau  c_{7}^{\tau^\prime }  \right]
       + {J_\chi (J_\chi+1) \over 12} \left[ c_4^\tau c_4^{\tau^\prime} +  \right.\nonumber \\
       &&\left. {q^2 \over m_N^2} c_9^\tau c_9^{\tau^\prime }+{v_T^{\perp 2} \over 2} \left(c_{12}^\tau-{q^2 \over m_N^2}c_{15}^\tau \right) \left( c_{12}^{\tau^\prime }-{q^2 \over m_N^2}c_{15}^{\tau \prime} \right) +{q^2 \over 2 m_N^2} v_T^{\perp 2}  c_{14}^\tau c_{14}^{\tau^\prime } \right] \nonumber \\
     R_{\Delta}^{\tau \tau^\prime}\left(v_T^{\perp 2}, {q^2 \over m_N^2}\right)&=&  {J_\chi (J_\chi+1) \over 3} \left[ {q^2 \over m_N^2} c_{5}^\tau c_{5}^{\tau^\prime }+ c_{8}^\tau c_{8}^{\tau^\prime } \right] \nonumber \\
 R_{\Delta \Sigma^\prime}^{\tau \tau^\prime}\left(v_T^{\perp 2}, {q^2 \over m_N^2}\right)&=& {J_\chi (J_\chi+1) \over 3} \left[c_{5}^\tau c_{4}^{\tau^\prime }-c_8^\tau c_9^{\tau^\prime} \right].
 \label{eq:R}
\end{eqnarray}


\providecommand{\href}[2]{#2}\begingroup\raggedright\endgroup

\end{document}